\documentclass[aps,prl,reprint,twocolumn,superscriptaddress,floatfix,nofootinbib, nobibnotes]{revtex4-2}
\usepackage{epsfig,amsmath,amssymb,color,comment,physics, dsfont}
\usepackage[makeroom]{cancel}
\usepackage[caption=false]{subfig}
\usepackage{mathrsfs}
\usepackage[countmax]{subfloat}
\usepackage[normalem]{ulem}
\usepackage[english]{babel}
\usepackage{dsfont}
\usepackage{float}
\usepackage[dvipsnames]{xcolor}
\usepackage{tikz}
\usepackage{nicefrac}
\usepackage{blindtext}
\usepackage{transparent}
\usepackage{bibentry}
\usepackage{titletoc}

\usepackage[bookmarks=true,colorlinks,linkcolor=RoyalBlue,urlcolor=RoyalBlue,citecolor=RoyalBlue]{hyperref}
\usepackage{orcidlink}

\newcommand{\sx}{\hat{\sigma}^x}
\newcommand{\sy}{\hat{\sigma}^y}
\newcommand{\sz}{\hat{\sigma}^z}
\newcommand{\spl}{\hat{\sigma}^+}
\newcommand{\smi}{\hat{\sigma}^-}
\newcommand{\spm}{\hat{\sigma}^\pm}

\newcommand{\T}[2]{\hat{T}_{#1 ; #2}}
\newcommand{\Tdag}[2]{\hat{T}_{#1 ; #2}^{\dag}}
\newcommand{\Hq}{\hat{H}_{q}}

\newcommand{\zp}[1]{#1}

\graphicspath{{./}}
\contentsmargin[2em]{1em}
\dottedcontents{section}[0em]{}{0em}{0.3pc}

\begin{document}

\title{
Hilbert space fragmentation at the origin of disorder-free localization\texorpdfstring{\\}{} in the lattice Schwinger model
}

\author{Jared Jeyaretnam \orcidlink{0000-0002-8316-9025}}\thanks{These authors contributed equally}
\affiliation{School of Physics and Astronomy, University of Leeds, Leeds LS2 9JT, UK}

\author{Tanmay Bhore \orcidlink{0000-0002-9304-7144}}\thanks{These authors contributed equally}
\affiliation{School of Physics and Astronomy, University of Leeds, Leeds LS2 9JT, UK}

\author{Jesse J. Osborne \orcidlink{0000-0003-0415-0690}}
\affiliation{School of Mathematics and Physics, The University of Queensland, St. Lucia, QLD 4072, Australia}

\author{Jad C. Halimeh \orcidlink{0000-0002-0659-7990}}
\affiliation{Max Planck Institute of Quantum Optics, 85748 Garching, Germany}
\affiliation{Department of Physics and Arnold Sommerfeld Center for Theoretical Physics (ASC), Ludwig Maximilian University of Munich, 80333 Munich, Germany}
\affiliation{Munich Center for Quantum Science and Technology (MCQST), 80799 Munich, Germany}

\author{Zlatko Papi\'c \orcidlink{0000-0002-8451-2235}}
\affiliation{School of Physics and Astronomy, University of Leeds, Leeds LS2 9JT, UK}

\date{\today}
\begin{abstract}
    Lattice gauge theories, discretized cousins of continuum gauge theories arising in the Standard Model, have become important platforms for exploring non-equilibrium quantum phenomena.
    Recent works have reported the possibility of disorder-free localization in the lattice Schwinger model.
    Using degenerate perturbation theory and numerical simulations based on exact diagonalization and matrix product states, we perform a detailed characterization of thermalization breakdown in the Schwinger model, including its spectral properties, structure of eigenstates, and out-of-equilibrium quench dynamics.
    We scrutinize the strong-coupling limit of the model, in which an intriguing double-logarithmic-in-time growth of entanglement was previously proposed from the initial vacuum state.
    We identify the origin of this ultraslow growth of entanglement as due to approximate Hilbert space fragmentation and the emergence of a dynamical constraint on particle hopping, which gives rise to sharp jumps in the entanglement entropy dynamics within individual background charge sectors.
    Based on the statistics of jump times, we argue that entanglement growth, averaged over charge sectors, is more naturally explained as either single-logarithmic or a weak power law in time.
    Our results suggest the existence of a single ergodicity-breaking regime due to Hilbert space fragmentation, whose properties are reminiscent of conventional many-body localization within numerically accessible system sizes. 
\end{abstract}
\maketitle

Lattice gauge theories (LGTs), originally introduced as powerful approximations to the continuous gauge theories underlying the Standard Model~\cite{Wilson1974,Kogut_Susskind_LGT_PRD_1975,montvay1994quantum}, have recently generated a flurry of interest for their realizations in synthetic quantum systems~\cite{Banuls2020,Bauer2023,dimeglio2023quantumcomputinghighenergyphysics,halimeh2023coldatom}. Gauge invariance endows LGTs with superselection sectors, determined through Gauss's law by the joint configuration of fermions and gauge fields. In one spatial dimension, it is possible to integrate out the gauge fields, which leaves the matter fields in the presence of an effective local potential furnished by the background charges, i.e., the eigenvalues of the Gauss law generator. The presence of dynamical constraints induced by the Gauss law make LGTs an attractive platform for exploring various interaction-driven forms of ergodicity breaking, such as many-body localization (MBL)~\cite{Basko2006,Gornyi2005,NandkishoreHuse,AbaninRMP,AbaninPapicAnnals,AletReview,SierantReview}, Hilbert space fragmentation~\cite{Sala2020,Khemani2020,MoudgalyaFragmentation}, and quantum many-body scars~\cite{Serbyn2021Review, MoudgalyaReview, ChandranReview}. 

The properties of MBL systems, such as suppressed transport and memory of initial conditions, continue to attract much attention in experiment~\cite{Schreiber2005,Choi2016,Bordia2016,Smith2016,Roushan2017,Rispoli2019,LukinMBL2019, Leonard2023, Yao2023,Stanley2023}.
The onset of MBL has been phenomenologically explained by ``local integrals of motion'' (LIOMs)~\cite{Serbyn13-1,HuseLIOMs,Ros2015}, an extensive set of conserved quasilocal operators whose eigenvalues fully characterize the eigenstates of MBL systems and thereby cause a breakdown of thermalization.
The LIOM picture is supported by analytical results for a specific model~\cite{Imbrie2014} and numerical simulations of a wider family of MBL models~\cite{PalHuse,Serbyn13-1,Kjall2014,Luitz2015,BarLev2015,Mondaini2015,Chandran2015,OBrien2016,Rademaker2016,Gray2018,Mierzejewski2018,AbaninChallenges, Theveniaut2020, Pietracaprina2021, Morningstar2022, Jeyaretnam2023}.
Nevertheless, strong finite-size effects often encountered in the numerics~\cite{Panda2019,Sieran2022} have recently raised questions about the stability of LIOMs in the thermodynamic limit~\cite{Suntajs2020,SelsPolkovnikov2021}. 

The standard models of MBL typically involve a static external potential that takes random values and acts as quenched disorder. By contrast, in LGTs, averaging over superselection sectors assumes the role of an effective disorder average for the matter field. The resulting \emph{disorder-free localization}~\cite{Smith2017DFL,Smith2017, brenes2017many,Metavitsiadis2017thermal,Nandkishore2017,Smith2018dynamical,Akhtar2018,Park2019glassy,Chanda2020,Russomanno2020homogeneous,Papaefstathiou2020dfl,McClarty2020dfl,Hart2021logarithmic,Zhu2021subdiffusive,Karpov2021dfl,Sous2021dfl,Halimeh2022enhancing,Chakraborty2022dfl,halimeh2022temperature,osborne2023dfl,Gao2023,Sala2024dfl,Hu2024} has been observed by preparing the system in a far-from-equilibrium initial state that is in a superposition of an extensive number of superselection sectors.
This makes LGTs reminiscent of other types of translation-invariant lattice models that had been proposed to self-consistently undergo an MBL transition due to interactions or dynamical constraints~\cite{Grover2014,Andraschko2014,Veness2017,Garrison2017,YaoDisorderFreeMBL, Schiulaz2015, Papic2015, vanHorseen2015, Hickey2016, Enss2017, vanNiewenburg2019, Schulz2019, Taylor2020, Kloss2023}.

A salient question is whether the nature of ergodicity breaking in LGTs is indeed of the same kind as in disordered MBL models~\cite{PalHuse}. For one, the effective disorder in LGTs is discrete due to the integer-valued background charges. Moreover, in a $U(1)$ LGT quenched from the fermonic vacuum state, the growth of entanglement was found to be ultraslow, seemingly following an unusual \textit{double-logarithmic} dependence on time~\cite{brenes2017many}, which was also reported in a disordered Bose-Hubbard chain~\cite{Chen2023b}. 
Such ultraslow entanglement growth appears distinct from a logarithmic-in-time entanglement growth in conventional MBL systems quenched from product states~\cite{Znidaric2008, Bardarson2012, Serbyn13-2}.  While the latter is a natural consequence of the LIOM-induced dephasing processes~\cite{Serbyn13-1}, to date there has been no understanding of the putative double-logarithmic growth of entanglement in a $U(1)$ LGT.  
  
In this paper, we study ergodicity breaking in a $U(1)$ LGT -- the one-dimensional (1D) Schwinger model -- using a combination of analytical and numerical tools, including degenerate perturbation theory, exact diagonalization and matrix-product states.
We characterize the model using both spectral and eigenstate properties, focusing on entanglement in particular, and we contrast the results against conventional MBL in disordered spin chains.
We find that the strong-coupling limit of the theory is dominated by an approximate Hilbert space fragmentation~\cite{Mallick_fragmentation}, which strongly impacts the properties of finite-size systems, causing visible deviations of the level statistics from standard random-matrix-theory ensembles.
We identify the mechanism of entanglement growth as sharp jumps between the weakly-connected Krylov sectors of the Hilbert space for a given charge sector of the gauge field.
The averaging over sectors reproduces the slow growth of entanglement previously observed~\cite{brenes2017many}; however, a closer look at the statistics of jump times suggests that the growth can be more simply explained as a weak power law or (single) logarithmic dependence on time. Our results point to  the existence of a single ergodicity-breaking regime in the lattice Schwinger model, which originates from Hilbert space fragmentation but mimics conventional MBL in numerically-accessible system sizes.

\section{Results}

\subsection{The dynamical phase diagram of the lattice Schwinger model}\label{sec:phase_diagram}
\begin{figure}[tbp]
    \includegraphics[width=\linewidth]{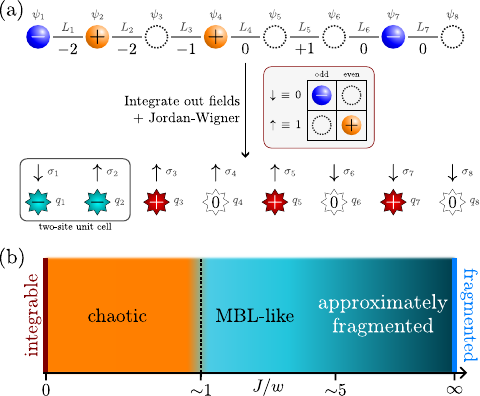}
    \caption{%
    	(a)~The 1D lattice Schwinger model (top) describes quantum electrodynamics on a lattice.
	    Fermionic degrees of freedom $\psi_n$ on odd (even) sites represent the presence or absence of an electron (positron), while the electric fields $L_n$ reside on the bonds and mediate coupling between the particles.
	    In 1D, one can integrate out the fields and apply a Jordan-Wigner transform, which results in a an XY-type model for spins $\sigma=\uparrow,\downarrow$.
	    The dynamical fields are replaced by static background charges $q_n$, which act as a disorder potential for spins.\linebreak[2]
	    (b)~Schematic phase diagram for the 1D lattice Schwinger model as a function of the dimensionless coupling ratio $J/w$ [see Eq.~\eqref{eq:spin_ham} for details].
	    In the weak-coupling limit $J/w \to 0$, the model reduces to an integrable XY spin chain.
	    In the thermodynamic limit, any finite $J$ breaks integrability, resulting in a chaotic phase.
	    Around $J/w \sim 1$, an ergodicity-breaking transition, consistent with an onset of MBL, has previously been observed~\cite{brenes2017many}.  However, the nature and extent of MBL phase is difficult to ascertain in finite-size systems due to its proximity to the regime dominated by Hilbert space fragmentation. The latter is exact at $J/w \to\infty$ and, as shown in this paper, strongly affects the properties of numerically-accessible systems even at \zp{$J/w \sim 5$}.     
    }
    \label{fig:intro_diagram}
\end{figure}

\zp{
The Schwinger model describes quantum electrodynamics on a one-dimensional (1D) lattice~\cite{Schwinger_QED_PRE}.
In the Wilson formulation~\cite{Wilson1974}, the model describes the coupling between matter fields that reside on lattice sites and $U(1)$ gauge fields that reside on the links between the sites, see Fig.~\ref{fig:intro_diagram}(a). The model is described by the Kogut-Susskind Hamiltonian~\cite{Kogut_Susskind_LGT_PRD_1975} 
\begin{equation}\label{eq:fermion_ham}
	\begin{aligned}
		\hat{H}_\text{Sch} &= -iw \sum_{n=1}^{N-1}[\hat{\Psi}_n^{\dag} \hat{U}_n \hat{\Psi}_{n+1} - \text{H.c.}] \\
		&+ J\sum_{n=1}^{N-1} \left(\hat{L}_n + \frac{\theta}{2\pi} \right)^2 +m\sum_{n=1}^N(-1)^n\hat{\Psi}_n^{\dag} \hat{\Psi}_n\ ,
	\end{aligned}
\end{equation}
where $\hat{\Psi}_n, \hat{\Psi}_n^{\dag}$ are the fermionic annihilation and creation operators on the $n$th lattice site, while $\hat{U}_n = e^{i \hat{\phi}_n}$ are the $U(1)$ parallel transporters defined on the bond between sites $n$ and $n{+}1$, see Fig.~\ref{fig:intro_diagram}(a).
Each $\hat{U}_n$ has a corresponding electric field operator $\hat{L}_n = -i \partial/\partial \hat{\phi}_n$, such that the commutation relation $[\hat{L}_n, \hat{U}_n] = \hat{U}_n$ holds.
$\theta$ describes a constant classical background field and can be used to tune between the confined ($\abs{\theta} < \pi$), and deconfined ($\abs{\theta} = \pi$) phases.
Throughout this paper, we will assume open boundary conditions (OBCs). A self-contained derivation of the LGT model~\eqref{eq:fermion_ham} from the continuum theory is provided in the Supplementary Information (SI)~\cite{SupplementaryInformation}.
}

\zp{
Physical states of the lattice gauge theory are constrained by Gauss' law, which is encoded as a set of local constraints on the lattice.
Specifically, to ensure gauge invariance, we consider the generators of the Gauss law,
\begin{equation}
 \hat{G}_n = \hat{L}_n - \hat{L}_{n-1} - \Psi_n^{\dag}\Psi_n + \frac{1}{2}[1 - (-1)^n]\ ,   
\end{equation}
such that $[\hat{H}_\text{Sch}, \hat{G}_n] = 0$. 
Physical states are defined as the eigenstates of $\hat{G}_n$, i.e., $\hat{G}_n \ket{ \Psi_{ \left\{q_\alpha\right\} } } = q_n \ket{ \Psi_{ \left\{q_\alpha\right\} } }$, 
where the eigenvalues $\left\{q_\alpha\right\}$ and corresponding states $\ket{ \Psi_{ \left\{q_\alpha\right\} } }$ define the background charge sector of the Hilbert space.
}

After a Jordan-Wigner transformation~\cite{SupplementaryInformation}, Gauss' law allows us to sequentially integrate out the gauge fields, resulting in an effective spin-1/2 Hamiltonian \zp{[see Fig.~\ref{fig:intro_diagram}(a)]}:
\begin{equation}\label{eq:spin_ham}
\hat{H}_{\left\{q_\alpha\right\}} = \hat{H}_{\pm} + \hat{H}_{ZZ} + \Hq,
\end{equation}
which explicitly depends on the background charge sector $\{q_\alpha\}$. The three terms are given by
\begin{align}
    \label{eq:spin_ham_XY}
    \hat{H}_{\pm} &= w \sum_{j=1}^{N-1}\left[\spl_j \smi_{j+1} + \text{H.c.}\right]\ , \\
    \label{eq:spin_ham_ZZ}
    \hat{H}_{ZZ} &= \frac{J}{2} \sum_{j=1}^{N-2} \sum_{k = j+1}^{N - 1}(N - k) \sz_j \sz_k\ ,\\
    \label{eq:spin_ham_charge_background} 
    \Hq &= \sum_{k=1}^{N} \left(h_k + \frac{m}{2} (-1)^k\right) \sz_k\ , \\
    \label{eq:spin_ham_local_fields}
    h_k &= \frac{J}{2} \Biggl(\zp{(N - k)\frac{\theta}{\pi}} -\left\lceil \frac{N - k}{2} \right\rceil + \zp{2} \sum_{j = k}^{N - 1} \sum_{i = 1}^{j} q_i \Biggr) \ ,
\end{align}
where $\spm \equiv (\sx \pm i \sy)/2$ denote the standard Pauli raising and lowering operators, and $\lceil \cdots \rceil$ is the ceiling function.
The three terms in the Hamiltonian correspond, respectively, to the number-conserving hopping of fermions across the bonds (the $XY$ spin term), the interactions, and the effective local field generated by the background charges $\{q_{\alpha}\}$. 
Note the asymmetric nature of long-ranged couplings in $\hat{H}_{ZZ}$: each spin interacts with all spins to its left with a constant strength, while the strength of interaction decreases linearly with distance for all spins to its right.
\zp{The Hamiltonian has a $U(1)$ magnetization-conservation symmetry, equivalent to charge conservation in the fermionic LGT, and here onward we restrict the Hamiltonian to the largest symmetry sector with zero total magnetization (i.e.\ zero charge).}
For simplicity, we focus on the massless case, $m=0$, and we express the hopping amplitude $w$ and interaction strength $J$ in units $\hbar=1$;
\zp{however, we expect the phenomenology to be broadly similar for any $m \ll J$.}
\zp{We also normalize the Hamiltonian by picking $w = 1$.
Finally, we choose $\theta = \pi$, placing the model in the deconfined phase, in order to rule out confinement as an explanation for the observed phenomena.}

The full state of the system encodes the degrees of freedom of both fermions and gauge fields: it is spanned by tensor products, $\ket{\Psi}_0 = \ket{\Psi}_f \otimes \ket{\Psi}_g$,
where $\ket{\Psi }_{f/g}$ is the state of the fermions/gauge fields. 
For the fermions, we focus on the  vacuum state or the N\'eel state of fermions, $\ket{\Psi}_f = \ket{101010\dots} \equiv  \ket{\text{vac}}_f$,
while other choices of fermionic states are discussed in SI~\cite{SupplementaryInformation}.
For $\ket{\Psi}_g$, following Ref.~\cite{brenes2017many}, we consider a uniform superposition of three eigenstates of the electric field operators -- see Methods. Due to the constraint between the gauge fields $\hat{L}_n$ and the background charges $q_n$, our initial state therefore effectively encodes a superposition over the charge sectors. 

The presence of disconnected charge sectors in the model, each evolving independently in time, produces an effective disorder landscape: each charge sector acts as a disorder realization, whose distribution is set by the full initial state of the fermions and gauge fields. This gives rise to a dynamical phase diagram sketched in Fig.~\ref{fig:intro_diagram}(b). At $J/w=0$, the model reduces to an integrable $XY$ spin chain. Finite values of $J/w\lesssim 1$ make the dynamics chaotic. Increasing the coupling further to  $J/w \gtrsim 1$ breaks ergodicity and was previously suggested to give rise to disorder-free localization~\cite{ brenes2017many,Giudici_MBL_U1}.
Treating the gauge field as a spin-1/2 degree of freedom, similar dynamics was found after including the four-fermion interaction term~\cite{Gao2023}.
On the other hand, the entanglement entropy growth at $J/w=10$~\cite{brenes2017many} suggested a much slower and parametrically different growth compared to the conventional LIOM picture of MBL, raising the possibility of a distinct MBL-like regime in the strong-coupling limit of the Schwinger model. Below we provide a detailed characterization of the phase diagram in Fig.~\ref{fig:intro_diagram}(b), focusing on the strong-coupling regime.  

\subsection{Level statistics and the density of states}\label{sec:spectral}

A standard metric of quantum chaos is the level spacing ratio~\cite{Oganesyan_Huse_PRB_2007}, $r = \min\{s_n, s_{n+1}\}/ \max\{ s_{n}, s_{n+1}\}$, which characterizes the spacing of adjacent energy levels $s_n = E_{n+1} - E_n$.
After all symmetries are resolved, the averaged energy gaps of an integrable or MBL system are expected to follow the Poisson distribution with the average ratio $\langle r \rangle_P \approx 0.386$, while those of chaotic systems with time-reversal symmetry obey the Gaussian Orthogonal Ensemble with $\langle r \rangle_\text{GOE} \approx 0.536$~\cite{Mehta2004}. 
We note that the spin Hamiltonian in \eqref{eq:spin_ham} is real in the computational basis and thus time-reversal symmetric, hence its ergodic phase should belong to the GOE class.

\begin{figure}[tbp]
	\includegraphics[width=\linewidth]{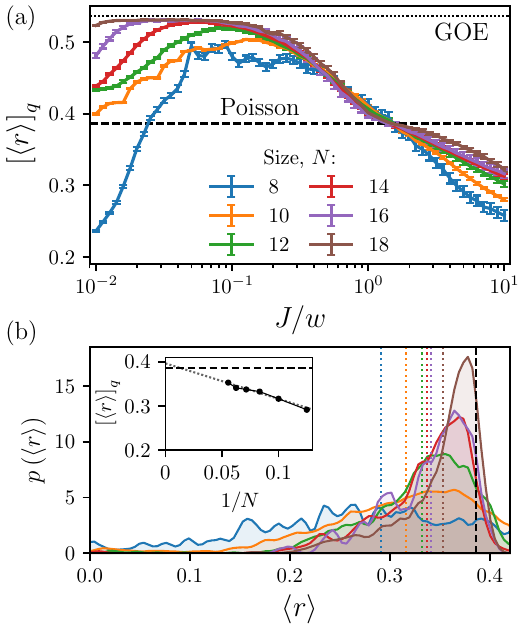}
	\caption{%
		(a) $\left\{q_\alpha\right\}$-averaged level spacing ratio as a function of the coupling strength parameter $J/w$, for 1000 background charge sectors.
		We truncate each spectrum to the $k = \min(\zp{1000}, \mathcal{N}/3)$ states with energies closest to the peak in the density of states, or an approximation thereof for $N = 18$, where $\mathcal{N}$ is the Hilbert space dimension.
		We also show the Poisson and GOE values as dashed and dotted black lines, respectively.
		The narrow MBL-like plateau appears at around \zp{$J/w \sim 1.5$}.
		(b) Distribution of the level spacing ratio across different charge sectors for fixed \zp{$J/w = 5$}.
		The average values are also shown as vertical dotted lines, and the black dashed line indicates the Poisson value.
		While the distributions clearly shift towards Poisson as $N$ increases, there is considerable weight at sub-Poisson values.
		The inset shows the trend in the $\left\{q_\alpha\right\}$-averaged value with $1/N$, showing that it likely reaches Poisson in the thermodynamic limit.
	}
	\label{fig:level_stats}
\end{figure}

In Fig.~\ref{fig:level_stats}(a), we calculate the sector-averaged level spacing ratio $\left[\langle r \rangle\right]_q$ as a function of the coupling strength $J$.
We consider system sizes up to $N = 18$ and 1000 randomly chosen background charge sectors.
Here, $\left[\dots\right]_q$ indicates an average over charge sectors, while an average over eigenstates is specified by $\langle \dots \rangle$.
For $N \leq 16$, we calculate the full spectrum and truncate to the $k$ states closest to the peak in the density of states (DOS), with $k$ the smaller of 500 or one-third of the Hilbert space dimension.
For $N = 18$, we use the shift-invert algorithm \cite{SLEPc} to obtain 500 eigenvalues closest to a target energy: we choose this target to be the modal classical energy (diagonal matrix element of the Hamiltonian), which is taken as an approximation to the DOS peak.

To understand Fig.~\ref{fig:level_stats}(a), we first note that the model is integrable for $J = 0$~\cite{sutherland2004beautiful}, with the average level spacing ratio equal to the Poisson value.
Upon introducing a non-zero but small $J$, integrability is broken, and we observe that $\left[\langle r \rangle\right]_q$ increases away from the Poisson value, steadily approaching the GOE value for a broad range of coupling strengths below \zp{$J \approx 0.1$}.
This is observed for all but the smallest system sizes $(N = 8, 10)$, suggesting that finite-size effects in our model are considerably stronger than in disordered MBL models~\cite{PalHuse}.
In the regime \zp{$J\lesssim 0.1$}, we expect the system to obey the Eigenstate Thermalization Hypothesis (ETH), with various consequences for the dynamics, such as relaxation of local observables to their thermal values, which were observed in Ref.~\cite{brenes2017many} for $J/w = 0.1$.

Upon increasing $J/w$ beyond \zp{0.2}, we see a deviation from GOE statistics with $\left[\langle r \rangle\right]_q$ steadily decreasing until a plateau forms at the Poisson value at around \zp{$J/w = 1.5$}.
This transition from GOE to Poisson statistics marks the onset of an MBL-like \zp{regime} at intermediate coupling strengths.
However, upon further increasing $J/w$, we observe a further decrease of $\left[\langle r \rangle\right]_q$, even dropping \emph{below} the Poisson value for all system sizes considered. 
This indicates the level statistics of the system is no longer well described by one of the standard random-matrix theory ensembles.
Such a dip is caused by the presence of a large number of (near)-degeneracies in the spectrum, the expected source of which could be an unresolved or an emergent symmetry.
However, we note that model hosts no such extra symmetry, even within the zero-magnetization sector.
In  Fig.~\ref{fig:level_stats}(b), we show the distribution of $\langle r \rangle$ -- that is, the average level spacing ratio within individual charge sectors -- across different charge sectors, for various system sizes at \zp{$J/w = 5$}.
The distribution becomes more sharply peaked and moves towards the Poisson value with increasing system size;
plotting the sector average against $1/N$, as in the inset, shows that it is likely the level spacing ratio will attain the Poisson value in the thermodynamic limit $N \to \infty$.  

\begin{figure}[tbp]
    \includegraphics[width=\linewidth]{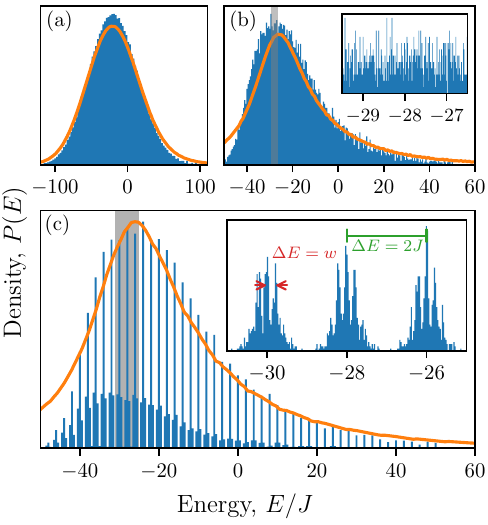}
    \caption{%
        Density of states  (DOS) for the charge sector
        \zp{$q_\alpha = \left[1\,\overline{1}\,0\,0\,1\,\overline{1}\,1\,0\,0\,1\,0\,0\,0\,\overline{1}\,\overline{1}\,0\,0\,0\right]$},
        where $\overline{1} \equiv -1$, and (a)~$J = 0.1$, (b)~$J = 1$, and (c)~\zp{$J = 5$}.
        All data is for system size $N=16$ and $w=1$.
        We also show, in orange, the averaged DOS across 250 superselection sectors.
        For (b) and (c), the insets show a close-up around the DOS peak.
        while the overall DOS looks similar for $J = 1$ and \zp{$J = 5$}, we see that for \zp{$J = 5$} the spectrum has split into well-separated ``towers'' separated by intervals of \zp{$2J$}.
        These towers are themselves split into peaks with interval of $w$.
    }
    \label{fig:dos}
\end{figure}
To understand the anomalous features in the level statistics, in Fig.~\ref{fig:dos} we plot the density of states (DOS) for three selected values of $J$ in a single charge sector at $N=16$.
We also show the averaged DOS across 250 charge sectors.
Furthermore, we choose to normalize the spectrum by dividing by $J$.
For small $J = 0.1$, in Fig.~\ref{fig:dos}(a), we find that the spectrum is relatively symmetric, with a peak close to $E = 0$ and a smooth Gaussian form consistent with typical chaotic systems.
However, for increasing $J$, the spectrum gains a large positive tail, while both the peak and the mean energy shifts towards the negative.
This skew is explained by the $\hat{H}_{ZZ}$ term in Eq.~\eqref{eq:spin_ham}: this is long-ranged, antiferromagnetic, and features a double-sum over terms with $\mathcal{O}(N)$-size coefficients, and therefore can give a large positive contribution up to $\mathcal{O}(JN^3)$ for certain states with many aligned spins.
On the other hand, the long-ranged antiferromagnetism leads to large frustration in trying to find the ground state, and empirically this will have an energy of $-\mathcal{O}(JN^2)$.

The dips in the level spacing ratio $\langle r \rangle$ are expected to be accompanied by sharp peaks in the DOS as energies cluster together to become almost degenerate.
Upon increasing the value of $J$, this is indeed what we observe as the DOS  becomes increasingly stratified, with sharp peaks clearly visible at \zp{$J/w = 5$} in Fig.~\ref{fig:dos}(c).
These sharp peaks are almost equally spaced in energy with a separation of \zp{$2J$}.
However, the slices used for calculating $\langle r \rangle$ in Fig.~\ref{fig:level_stats} should fit entirely within a single peak for the largest system sizes considered, and so this alone cannot explain the dip in $\langle r \rangle$ below $\langle r \rangle_\text{GOE}$.
Further zooming in on these peaks, we can see that they are themselves jagged and irregular, with principal peaks at intervals of $w$.
Therefore, not only are the strong-coupling terms $\hat{H}_{ZZ}$ and $\Hq$ restricting basis states with different classical energies from mixing effectively, these classical energy levels are also not being well-mixed by the hopping term $\hat{H}_\pm$.
This indicates that there is a dynamical restriction or Hilbert space fragmentation present in the model, which will be explored in detail below.

\subsection{Hilbert space fragmentation at strong coupling}\label{sec:fragmentation}

In the limit $J / w \to \infty$, the Hamiltonian~\eqref{eq:spin_ham} becomes diagonal and its eigenstates are simply product states in the $z$-basis.
We would like to understand the spectrum of the model as we reintroduce a perturbatively small $w$.
With $m = 0$, the coefficients in both $\hat{H}_{ZZ}$ and $\Hq$ take values which are multiples of $J/2$, while the constituent Pauli operators can only take the values $\pm 1$, thus it is clear that with $w = m = 0$, energies will take discrete values separated by (at least) $J$.
\zp{In fact, the separation is always at least $2J$ \cite{SupplementaryInformation}}.
The width of the spectrum will then be $\mathcal{O}(JN^3)$ (dominated by $\hat{H}_{ZZ}$), which implies there are at most $\mathcal{O}(N^3)$ of these energy levels.
With $2^N$ states in the Hilbert space, the eigenenergies will be massively degenerate in the thermodynamic limit.
We label these degenerate towers $\{\mathcal{K}_a\}$, $a \in \mathbb{Z}$.

As we perturbatively reintroduce $w$, we expect naively that the degenerate towers will broaden to a width $\mathcal{O}(Nw)$ due to $\hat{H}_{\pm}$.
However, we instead find that the structure of these towers, and the interplay between $\hat{H}_{ZZ}$ and $\Hq$, imposes a kinetic constraint which fractures the towers into disconnected subspaces.
It can be shown~\cite{SupplementaryInformation} that it costs zero energy to exchange two spins at sites $\ell$ and $\ell + 1$, $\ket{01} \leftrightarrow \ket{10}$, only if
%
%
\begin{equation}
    \label{eq:resonant_hop_condition}
    \zp{\sum_{j=1}^{\ell - 1} \sz_j + 2 \sum_{j=1}^\ell q_j = (\ell \!\!\! \mod 2) - \frac{\theta}{\pi}\ ,}
\end{equation}
while any other dynamics within a tower must occur as a second-order off-shell process involving another tower and is therefore suppressed by a factor of at least $\order{w/J}$.

\begin{figure}[tb]
	\includegraphics[width=1\linewidth]{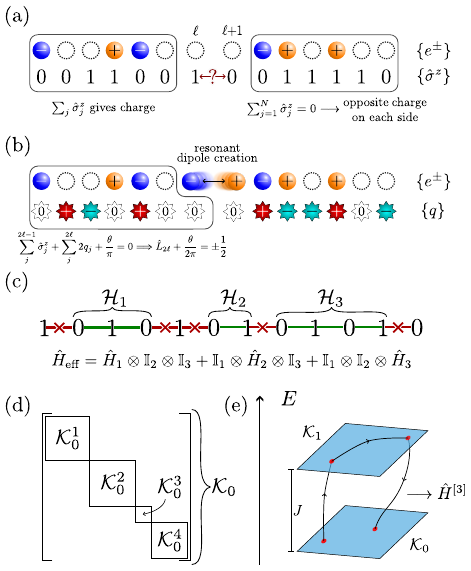}
	\caption{%
		(a)~The parity $\sum_j \sz_j$ over some region gives the total charge in the original gauge theory; this must sum to zero over the whole chain.\linebreak[2]
		(b)~Exchanging $\ket{10} \leftrightarrow \ket{01}$ at sites $2\ell$ and $2\ell + 1$ creates an $e^- e^+$ pair.
		\zp{This is resonant if the combination of gauge and background fields  $L_{2\ell} + \theta / 2 \pi = \pm 1/2$ both before and after this process.}\linebreak[2]
		(c)~For a particular charge sector and spin configuration, only certain bonds satisfy the previous condition and thus permit dynamics in the large-$J$ limit.
		These ``active regions'' are independent and so the effective Hamiltonian is a sum of local commuting terms.\linebreak[2]
		(d)~Each energy level $\mathcal{K}_a$ therefore fragments into Krylov subspaces $\{\mathcal{K}^b_a\}$ which are disconnected under the action of $\hat{H}_\pm$.\linebreak[2]
		(e)~For finite $J$, degenerate perturbation theory (see Methods) allows us to obtain an effective Hamiltonian within $\mathcal{K}_0$ as an expansion in powers of $w/J$.
		For example, a series of three spin exchanges via a nearby energy level contributes a term to the third-order effective Hamiltonian $H^{[3]}$ of magnitude $\order{w^3/J^2}$.
	}
	\label{fig:hopping_diagram}
\end{figure}

The background charges therefore set the condition for resonant hopping to occur.
For the initial state $\ket{\text{vac}}$, \zp{twice} the sum of the background charges is $\hat{L}_\ell(0)$, and so the initial electric field between two sites directly controls this hopping.
Physically, we can interpret this as the manifestation of the 1D Coulomb law.

More generally, $\sum_{j = 1}^{\ell - 1} \sz_j \! - \! (\ell \! \mod{2})$ counts the fermionic charge (that is, in terms of electrons $e^-$ and positrons $e^{+}$) to the left of the candidate spins, as shown in Fig.~\ref{fig:hopping_diagram}(a).
Since we have number conservation and half-filling, which is equivalent to enforcing charge neutrality, this quantity is equal and opposite on the other side of the bond.
Figure~\ref{fig:hopping_diagram}(b) then illustrates how exchange of spins results in the creation or annihilation of an $e^- e^+$ pair, \zp{increasing or decreasing $L_\ell$ by one.
If Eq.~\eqref{eq:resonant_hop_condition} is satisfied, then with $\theta = \pi$ this will take $L_\ell + \theta / 2\pi$ from $-1/2$ to $+1/2$ (or vice versa), such that the energy of the electric field $\propto (L_\ell + \theta / 2\pi)^2$ is unchanged.}
The net effect is that starting from a particular initial state, only certain parts of the chain -- ``active regions'' -- will permit dynamics in the large-$J$ limit [Fig.~\ref{fig:hopping_diagram}(c)].
We therefore see that each degenerate energy level $\mathcal{K}_a$ fractures into a set of subspaces $\{\mathcal{K}_a^b\}$ which are disconnected under the action of $\hat{H}_\pm$ [Fig.~\ref{fig:hopping_diagram}(d)].
Each of these subspaces is a Krylov subspace~\cite{MoudgalyaReview}, i.e., a space obtained by repeatedly applying an operator $ \hat{O}$ to an initial state $\ket{\psi}$, where, in this case, $\ket{\psi}$ is a $\sz$-basis state and $\hat{O}$ is the projection of $\hat{H}_\pm$ into a degenerate tower $\mathcal{K}_a$.

The fragmentation partially explains the behavior seen in the level statistics at large $J$ in Figs.~\ref{fig:level_stats}(a).
Because the resonance condition \eqref{eq:resonant_hop_condition} is invariant under rearrangements of spins that do not cross the bond under consideration, the active regions will have fully independent dynamics in the large-$J$ limit.
We can therefore typically decompose each Krylov subspace and its effective Hamiltonian (given by the projection of $H_\pm$),	$\mathcal{K} = \mathbb{C}^{n_1} \otimes \mathbb{C}^{n_2} \otimes \mathbb{C}^{n_3} \otimes \dots$ and 
	$\hat{H}_\mathcal{K} = \hat{H}_1 + \hat{H}_2 + \hat{H}_3 + \dots$, 
where the $n_i$ is the total Hilbert space dimension of the $i$th active region,
and $\hat{H}_i$ is the part of $\hat{H}_\pm$ acting only on $\mathbb{C}^{n_i}$.
The eigenvalues of $\hat{H}_\mathcal{K}$ will therefore be sums of individual eigenvalues, $E_\mathcal{K} = \varepsilon_1 + \varepsilon_2 + \varepsilon_3 + \dots$.
In general, the active regions and therefore the $n_i$ will be small, such that the $\varepsilon_i$ will be simple algebraic numbers (often integers, even zeros), leading to many accidental degeneracies in $E$ and thus bringing $\langle r \rangle$ below the Poisson value.
This also explains the spacing by $w$ observed in Fig.\ref{fig:dos}(c).

As $N \gg 1$, the active regions will typically remain small, but increase in number, such that the distribution of eigenvalues $\sum_j \varepsilon_j$ would begin to resemble something closer to a Poisson distribution.
The width of this distribution would also grow as $\mathcal{O}(\sqrt{N})$, weakening the strong separation in energy scales and therefore increasing mixing between states belonging to each tower.
Fig.~\ref{fig:level_stats}(b) indeed shows that that the distribution of the level spacing ratio tends towards the Poisson value as we keep $J/w$ fixed and increase $N$.

While the Hilbert space fragmentation discussed above is only strictly valid for $J / w \to \infty$, the structure and dynamics of the model for large but finite $J / w$ can be captured using degenerate perturbation theory (DPT), by which the effective Hamiltonian within a $J$-tower $\mathcal{K}_0$ is expanded in orders of $w / J$, $\hat{H}_\text{eff} = \hat{H}^{[0]} + \hat{H}^{[1]} + \hat{H}^{[2]} + \dots$, with $\hat{H}^{[n]} = \order{w^n/J^{n-1}}$, see Methods. This means that evolution under the Hamiltonian up to $n$-th order, which we label $\mathrm{DPT}(n)$, will typically capture the dynamics up to $wt = \order{(J/w)^{n-1}}$, so long as $J/w \gg 1$, and will involve processes with $n$ virtual hops via $n-1$ (not necessarily unique) energy levels.
An illustration of a third-order process is given in Fig.~\ref{fig:hopping_diagram}(e).
In the Methods, we demonstrate that third-order DPT accurately captures the dynamics of local observables and entanglement growth during intermediate times $w t\lesssim 10^3$.
However, to capture the putative double-logarithmic entanglement growth regime, discussed next, one would have to extend DPT to much higher orders, which was not practically feasible. 

\begin{figure*}
	\includegraphics[width=\linewidth]{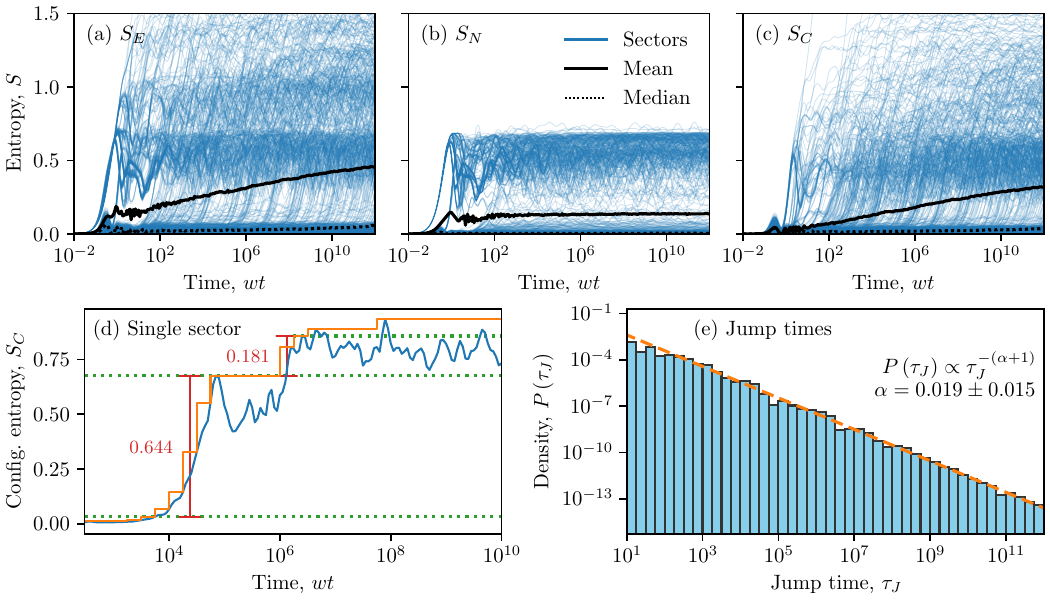}
	\caption{%
        (a-c) Entanglement entropy $S_E$, number entropy $S_N$, and configurational entropy $S_C$,  respectively [Eqs.~\eqref{eq:entropy_decomposition}-\eqref{eq:config_entropy}], for $J=5$ and $N = 16$.
        In each figure we show the individual results for each of \zp{1000} charge sectors (blue), as well as the mean (black solid) and median (black dotted) averages over sectors.
        Jumps in the configurational entropy are clearly visible across a wide range of time scales, spanning many orders of magnitude, with a single exemplary charge sector shown in (d). 
        By taking the cumulative maximum $\sup_{\tau < t_j} S(\tau)$ of the entropy growth at discrete times $t_j$, thus approximating it as a series of steps (orange), we can calculate the heights and times $\tau_J$ of the jumps (red).
        (e) Histogram of jump times $\left\{\tau_J\right\}$, weighted by jump height.
        A clear fit to a power law (orange dashed line) is observed, with $P(\tau_J) \propto \tau_J^{-(\alpha + 1)}$ and \zp{$\alpha = 0.019 \pm 0.015$}.
    }
    \label{fig:jump_times}
\end{figure*}
\subsection{Late-time entanglement growth}\label{sec:doublelog}
Entanglement provides a powerful diagnostic of thermalization and its breakdown. We characterize the entanglement of a pure state $\ket{\psi}$ via its entanglement entropy $S_E$ for a bipartition of the system into subsystems $A$ and $B$:
\begin{equation}
    \label{eq:entanglement_entropy}
    S_E = -\tr(\rho_A \ln \rho_A),\quad \rho_A = \tr_B \rho\ ,
\end{equation}
where $\tr_B$ is the trace over degrees of freedom in $B$, and $\rho = |\psi\rangle\langle\psi|$ is the density matrix for the full system. In chaotic systems evolving under unitary dynamics, $S_E$ increases as a power-law in time, $S_E(t)\sim t^\gamma$~\cite{KimHuse2013}. By contrast, in MBL systems the growth is logarithmic, $S_E(t)\sim \ln t$~\cite{Znidaric2008,Bardarson2012,Serbyn13-2}. Previous numerical simulations~\cite{brenes2017many} of the model in Eq.~\eqref{eq:spin_ham} argued for an even slower growth of entanglement entropy in the strong-coupling regime, conjecturing that it follows an unusual \textit{double}-logarithmic dependence $S_E(t) \sim \ln(\ln t)$ at late times. Here we explore the origin of such slow growth and its relation to Hilbert space fragmentation, focusing on individual charge sectors.

For short times $wt < J$ we expect the dynamics to remain entirely within the initial subspace $\mathcal{K}^0_0$, while at later times, higher-order processes allow the dynamics to escape both this subspace and the containing energy level.
However, we find that \zp{the proportion of the state which lies outside of the initial subspace} saturates by the time the proposed sub-logarithmic entanglement growth sets in.
We see similar trend in quantities which measure the distribution of coefficients across the computational ($\sz$) basis, such as the inverse participation ratio~\cite{Serbyn13-1}.
This implies that the ultraslow growth in entanglement following the initial transient is driven not by changes to the distribution of coefficients across the basis, but due to correlations built up between \textit{configurations} of particles either side of the bipartition.
From the numerical data, we also observe that $S_E(t)$ displays very different behaviors in individual charge sectors: while it quickly increases in some sectors, in others it remains close to zero for the entire accessible range.

In a particle-number conserving system, the reduced density matrix $\rho_A$,  Eq.~\eqref{eq:entanglement_entropy}, can be block-diagonalized, $\rho_A = \bigoplus_n p(n) \rho_A(n)$ with $\tr \rho_A(n) = 1$, where $\rho_A(n)$ corresponds to those states with $n$ particles in subsystem $A$, such that $p(n)$ is the probability distribution of particle number in $A$.
This allows us to decompose the total entropy into two contributions \cite{Kiefer-Emmanouilidis2021, Lukin2021}: the \textit{number} entropy $S_N$, which represents uncertainty in particle count, and the \textit{configurational} entropy $S_C$, which expresses our uncertainty in \textit{how} those fixed numbers of particles are arranged:
\begin{gather}
    \label{eq:entropy_decomposition}
	S_E = S_N + S_C\ ,\\
    \label{eq:number_entropy}
	S_N = -\sum_n p(n) \ln p(n)\ ,\\
    \label{eq:config_entropy}
	S_C = -\sum_n p(n) \tr[\rho_A(n) \ln \rho_A(n)] \ .
\end{gather}
Below we will argue that the origin of the ultraslow growth of entanglement are discrete \textit{jumps} in the configurational entropy $S_C$, which occur in particular charge sectors at well-defined times, and it is only after averaging over charge sectors that a smooth growth in the entropy emerges.
On the other hand, the number entropy $S_N$ saturates at relatively short times and so does not play a role in this slow growth.

These jumps correspond to the resolution of near-degeneracies that originate from the fractured nature of the Hilbert space.
\zp{In the large-$J$ limit, the unperturbed basis states are massively degenerate.
Each successive order $n$ of degenerate perturbation theory, which describes the system away from the limit, then provides a correction $\mathcal{O}(w^n/J^{n-1})$ to the energies, splitting these degeneracies.
The dynamics will resolve these corrections at a time $wt \approx (2J/w)^{n-1}$, resulting in a jump in the entanglement entropies as the corresponding states hybridize~\cite{Ghosh2022}}
 
In Figs.~\ref{fig:jump_times}(a)-(c) we show the entanglement, number, and configurational entropies, respectively, for the Schwinger model with $J = 8$, including both the results from each of 512 individual charge sectors as well as the mean and median over sectors.
Here, the median is defined as the median value at each moment in time -- it is important to note that, as a complete trajectory, this does not necessarily represent a single ``typical'' charge sector.
Throughout this section, we also convolve individual charge sectors with a narrow Gaussian window to eliminate high-frequency oscillations; this is not neccessary for the mean and median.

There are a few noteworthy features in Figs.~\ref{fig:jump_times}(a)-(c).
The most prominent is the very large spread in the entropies of different charge sectors: some of these stay close to zero, while some attain values two or three times larger than the mean.
This is especially true for the number entropy, $S_N$, which saturates quickly to a small value, with only certain charge sectors attaining values which cluster around ${\sim}\ln(2)$, which indicates the spread of a single particle across the boundary.
Furthermore, for the configurational entropy $S_C$, we observe that growth in individual charge sectors occurs via prominent jumps from one plateau to another (up to some weaker fluctuations around these plateaus).
The initial jumps from near-zero to values around $S_C \approx 0.5$ are visible from $t = 10^1$ all the way up to the latest times accessible, \zp{$t = 10^{12}$}.
\zp{We also observe that the median lags well behind the mean, staying close to zero at all numerically accessible times.}
Taken together, these features tell us that the slow growth in the entanglement entropy $S_E$ is driven not by steady growth across charge sectors, but by rapid jumps in the configurational entropy of individual sectors.
It is only when we average over the sectors that these jumps are smoothed out into the slow ``double-logarithmic'' growth found in Ref.~\cite{brenes2017many}.

We are therefore interested in characterizing the jumps in configurational entropy, as these could offer deeper insight into the functional form of the entanglement growth.
Figure~\ref{fig:jump_times}(d) shows the illustrative behavior of $S_C(t)$ for a single  charge sector.
By taking a cumulative maximum of $S_C(t)$ in a particular charge sector, i.e.,  $\sup_{\tau < t_j} S_C(\tau)$, at discrete time intervals $t_j$, we are able to locate intervals of growth in $S_C(t)$, followed by plateaus during which it does not exceed its historic maximum~\cite{Ghosh2022}.
Specifically, using intervals of $\Delta\log_{10}(t) = 0.25$, we consider periods where the cumulative maximum increases by at least \zp{$0.05$} in each consecutive interval.
We identify these periods as the jumps, calculating their height $\Delta S_C$ as the difference in $\sup_{\tau < t_j} S(\tau)$ between the beginning and end, and the jump time $\tau_J$ as the geometric midpoint of the start and end times.

\begin{figure}[tb]
    \includegraphics[width=\linewidth]{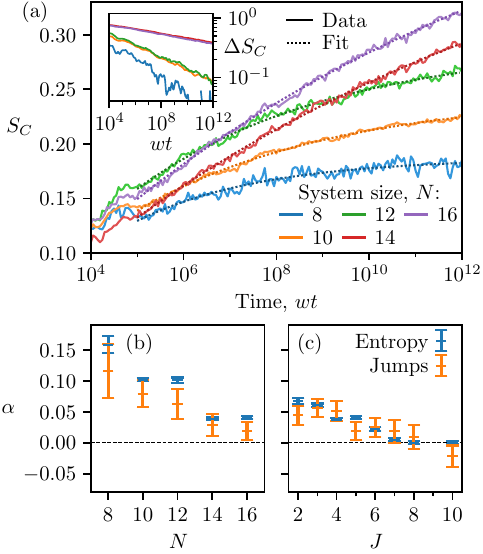}
    \caption{%
        (a) Sector-averaged configurational entropy, $S_C(t)$, for $J=5$ and system sizes $N = 8\text{--}16$ (solid lines).
        For each $N$ we fit a power-law of the form $S_C(t) = S_\infty - S_0 t^{-\alpha}$ (dotted lines).
        In the inset, we show the power-law decay towards the fitted saturation value, $\Delta S_C = S_\infty - S_C(t)$ for \zp{each $N$}.
        (b) The fitted exponent $\alpha$ as a function of $N$, calculated both from the entropy [as per panel (a)] (blue), or from the jump times [as per Fig.~\ref{fig:jump_times}(e)] (orange), for \zp{$J = 5$}.
        (c) The same as (b), but plotting the power-law exponent as a function of $J$ for fixed $N = 16$.
        We see good agreement between both methods of extracting $\alpha$.
    }
    \label{fig:jump_exponents}
\end{figure}

Finally, in Fig.~\ref{fig:jump_times}(e), we show the distribution of jump times $\tau_J$, where we have weighted the contribution of each jump to the histogram by the corresponding jump height $\Delta S_C$.
Once this distribution is computed, we fit it to a power law in time, $P(\tau_J) \propto \tau_J^{-(\alpha + 1)}$.
We note that, across various values of $N$ and $J$, the value of this exponent is not sensitive to the exact parameters used in the jump-finding algorithm, or even to the use of the jump heights as weightings, relative to the calculated errors.

The intuition here is that, if the entropy growth is driven by jumps at late times, then the integral of jump distribution will give back the configurational entropy, i.e., if $P(\tau_J) \propto \tau_J^{-(\alpha + 1)}$, then $S_C(t) = S_\infty - S_0 t^{-\alpha}$, with a power-law decay towards a steady-state saturation value $S_\infty$.
We confirm this intuition in Fig.~\ref{fig:jump_exponents}. 
Firstly, in Fig.~\ref{fig:jump_exponents}(a), we show the sector-averaged $S_C$ for $J = 8$ and several system sizes, all showing power-law decay towards a steady-state value, in agreement with the ansatz. Furthermore, in the inset of Fig.~\ref{fig:jump_exponents}(a), we show the difference between $S_C$ and $S_\infty$ on a log-log scale, verifying that these are indeed power-law decays.

We repeated the procedure above for a spread of different $N$ and $J$, calculating $\alpha$ both by characterizing the jump times and by fitting $S_C$ to a power law.
In Fig.~\ref{fig:jump_exponents}(b), for $J = 8$, we show $\alpha$ as a function of $N$, while in Fig.~\ref{fig:jump_exponents}(c), we plot $\alpha$ against $J$ for fixed $N = 16$.
These show a clear trend towards $\alpha = 0$ for increasing $N$ and $J$, which would \zp{be consistent with} logarithmic growth $S(t) \sim \ln(t)$.
\zp{We further see a trend of increasing saturation time for $S(t)$, which is evidence of unbounded entanglement growth in the thermodynamic limit.}
Taken together this suggests that, within the times accessible to double-precision numerics, the available data for entanglement growth at $J/w \gg 1$ is consistent with a \emph{single} logarithm and it is not necessary to invoke double-logarithmic scaling.

We note that our analysis above bears similarity with the one for the disordered XXZ model in Ref.~\cite{Ghosh2022}.
However, the latter model shows jumps in the \emph{number} entropy, which were used to model the double-logarithmic growth of particle fluctuations in the MBL phase.
In our case, the particle fluctuations appear strongly suppressed at all accessible timescales, and it is the \emph{configurational} entropy that undergoes jumps which give rise to the slow growth of entanglement.
We present a more detailed comparison between the $U(1)$ LGT and disordered XXZ models in the SI~\cite{SupplementaryInformation}.
In the SI, we also study a variant of the XXZ model with discrete disorder, for which the behavior of number and configurational entropies bear closer similarity with the $U(1)$ LGT, suggesting that the discrete nature of the disorder potential is indeed crucial for understanding the late-time dynamics of entanglement. 

\subsection{MBL-like regime at intermediate coupling}\label{sec:eigenstates}

\begin{figure}[tbp]
	\includegraphics[width=1\linewidth]{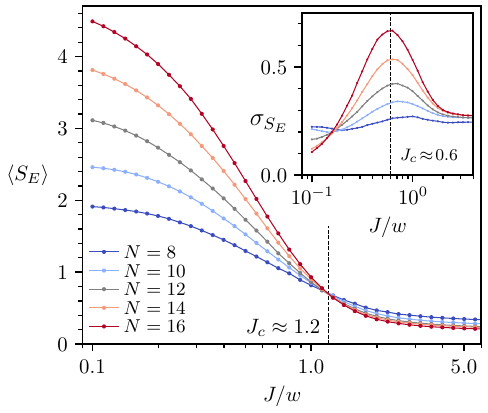}
	\caption{%
        The mean (main) and standard deviation (inset) of the bipartite  entanglement entropy of eigenstates, for 1/3 of eigenstates around the DOS peak in each charge sector.
        The data is further averaged over all possible sectors for $N = 8,10$; $1000$ sectors for $N = 12,14,16$; and $250$ charge sectors for $N = 16$.
        The dashed black lines mark the ``critical'' values of $J$.
        }
	\label{fig:eigenstate_entropies}
\end{figure}

Finally, we address the existence of MBL phase at intermediate couplings $J/w \sim 1$, where a Poisson plateau in the level statistics is clearly seen in Fig.~\ref{fig:level_stats}.
The pertinent question is whether this plateau should be interpreted as the system undergoing an MBL transition in the thermodynamic limit, or if it only arises because the accessible finite-size systems are still impacted by the residual effects of fragmentation at $J/w\to\infty$.
Our analysis of entropy jumps in Fig.~\ref{fig:jump_exponents} suggests that the exponent $\alpha$, although it remains small, steadily rises from zero as $J$ is reduced, suggesting glassy, power-law relaxation dynamics rather than localization.
\zp{On the other hand, we see $\alpha \to 0$ as $N$ increases, with an increasing saturation time, which is consistent with a logarithmic entanglement growth in the thermodynamic limit.}
However, at $J \sim w$, DPT is no longer valid and we do not expect to see the jumps in entanglement entropy observed at larger $J$, hence we do not expect Fig.~\ref{fig:jump_exponents} to accurately capture this regime. 

To probe the existence of MBL transition beyond the level statistics, we study the entanglement structure of eigenstates.
\zp{Close to the middle of the spectrum, the eigenstates of ergodic systems are well-modeled by featureless Haar-random vectors, thus their $S_E$ scales with the volume of the subsystem.
By contrast, mid-spectrum eigenstates in the MBL phase can be constructed through quasilocal unitary transformations from product states, hence their $S_E$ follows an area-law~\cite{Bauer2013}. To focus on mid-spectrum properties, we consider $33 \%$ of eigenstates around the DOS peak in each charge sector.}
In disordered spin chain models, a transition from volume-law to area-law entanglement entropy was observed by tuning the disorder strength, accompanied by a diverging variance of $S_E$ at the transition, even within a fixed disorder realization~\cite{Kjall2015}.
In finite-size numerics, the crossing point in the mean of $S_E$ of the eigenstates typically provides a lower bound for the critical disorder where the fluctuations of $S_E$ diverge, and the two estimates of the transition do not necessarily coincide in finite-size systems~\cite{SierantReview}. 

In Fig.~\ref{fig:eigenstate_entropies}, the ergodic nature of the system at small $J$ is witnessed by a volume-law scaling of the mean $S_E$  and the vanishing of its variance.
We calculate the mean and standard deviation of $S_E$ for 1/3 of eigenstates around the DOS peak within each charge sector, after which we average the data over charge sectors.
Upon increasing $J$, we observe the standard deviation peaks at \zp{$J_c \approx 0.6$}, while the mean entanglement entropies steadily decreases until a crossing occurs at \zp{$J_c \approx 1.2$}, consistent with an MBL transition. Upon further increasing $J$, the mean entanglement entropies become approximately independent of system size, as expected for area-law scaling.

Curiously, the variance of $S_E$ in Fig.~\ref{fig:eigenstate_entropies}(b) saturates to a non-zero value at \zp{large $J$}. A closer inspection of the distribution of $S_E$ reveals that most eigenstates have either zero or $\ln2$ entanglement entropy. This can be understood from the Krylov subspaces introduced above.  For sufficiently small $w/J$, the subspaces will not mix, and eigenstates may be decomposed into a product of states within each active region.
Let $p_\text{cross}$ be the proportion of basis states belonging to a Krylov subspace with an active region that crosses the central bond; only that same proportion $p_\text{cross}$ of eigenstates will have non-zero entanglement entropy for a cut through that bond.
If a single particle is delocalized across the boundary, the resultant entanglement entropy will be $\ln2$, and therefore the mean $\langle S_E \rangle \approx p_\text{cross} \ln2$ and the variance $\sigma_{S_E} \approx \sqrt{p_\text{cross}}\ln2$.
We find numerically that, for a random state in the half-filling sector and a random charge sector, the chance that condition~\eqref{eq:resonant_hop_condition} is satisfied for the central bond is $\mathcal{O}(N^{-1/2})$ for large $N$, in accordance with the probability that a random walk has zero displacement~\cite{weissteinRandomWalk}.
Therefore, we expect $\langle S_E \rangle \propto N^{-1/2}$ and $\sigma_{S_E} \propto N^{-1/4}$, \zp{which we observe when we do not restrict to the central one-third of the spectrum but include all eigenstates}.

\begin{figure}[tbp]
	\includegraphics[width=\linewidth]{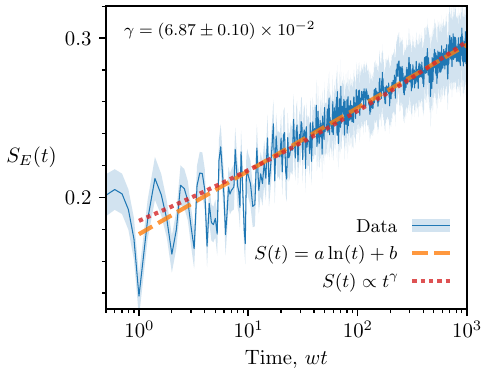}
	\caption{%
            Growth of entanglement entropy from the $\ket{\text{vac}}$ initial state for system size $N = 50$ at $J/w = 3$, obtained via MPS simulations (see Methods for details). The data is averaged over \zp{1000} background charge sectors, with the standard deviation represented by the shading. The dashed lines indicate different types of fits summarized in the legend: a logarithmic fit, and a power-law fit with exponent $\gamma = (6.87 \pm 0.10) \times 10^{-2}$.
	}
 \label{fig:entropygrowth}
\end{figure}       

One of the smoking-gun signatures of the MBL phase is the logarithmic in time growth of entanglement entropy from unentangled initial states~\cite{Znidaric2008,Bardarson2012}, a direct consequence of the the exponentially slow dephasing between LIOMs~\cite{Serbyn13-2}.
In disordered MBL models, this logarithmic growth of entanglement entropy typically persists over many decades in time, e.g., up to times as large as $10^{10}$ units in the system's natural inverse energy scale~\cite{Serbyn13-2}. 
Although in a finite system the entropy eventually saturates, the saturation time increases with system size, hence in the thermodynamic limit, the entropy growth is believed to be unbounded~\cite{Bardarson2012}. 

Entanglement entropy growth for the $U(1)$ LGT model at $J/w=3$ is presented in Fig.~\ref{fig:entropygrowth}. In small systems that can be studied by exact diagonalization, we find the universal regime of entropy growth to be quite short and impacted by the broad approach to the saturation value. Hence, in Fig.~\ref{fig:entropygrowth}, we used matrix product state (MPS) simulations in a large system of $N=50$ spins, as described in the Methods. 
This allows us to avoid the finite-size effects due to small chain sizes, however the times that can be reached are limited due to the increase in computational effort that comes from the build-up of entanglement. Within the available time range, entropy appears to follow $S_E\sim \ln(t)$ dependence, consistent with MBL. However, the accessible timescales are insufficient to reliably discriminate from power-law dependence, $S_E\sim t^\gamma$ (with $\gamma > 0$), as shown by the different fits in Fig.~\ref{fig:entropygrowth}. Thus, our data cannot rule out the possibility of slow, power-law delocalization at intermediate $J/w$ values. 

\section{Conclusions and discussion}\label{sec:conclusion}

We have performed a detailed characterization of ergodicity breaking in the lattice Schwinger model as a function of the coupling strength $J/w$. Standard metrics of quantum chaos, including the level statistics and eigenstate entanglement, with further results on the spectral form factor and many-body Thouless parameter in the SI~\cite{SupplementaryInformation}, all undergo a sharp change around $J/w \sim 1$. While these results are reminiscent of an emergent MBL phase, as also found in the \emph{bosonic} lattice Schwinger model~\cite{Chanda2020}, we have argued that the observed MBL-like signatures can be accounted for by the residual effects of Hilbert space fragmentation in the infinite-coupling limit. The fragmentation naturally follows from the discretized nature of the  model and should be present even in the original fermionic formulation, i.e., prior to substituting dynamical fields for static background charges.
\zp{Therefore, there are two limits of the model: if we take $J / w \to \infty$, the model remains fragmented even as $N \to \infty$.
However, if we take $N \to \infty$ first, at any finite $J/ w$ the towers at different energies will mix, and our evidence in Figs.~\ref{fig:level_stats} and~\ref{fig:jump_exponents} suggests this will approach Poisson level statistics and an MBL-like regime, although we cannot rule out eventual delocalization.}

Furthermore, we have identified the origin of the putative double-logarithmic entanglement growth in the strong-coupling regime~\cite{brenes2017many}, presenting its alternative interpretation as a slow approach to the steady state in finite-size systems. 
Importantly, this entropy growth was shown to occur largely via the rearrangement of particles in subsystems (the configurational entropy), rather than fluctuations in their numbers (the number entropy), as seen in disordered MBL models. We have also identified sharp jumps in the configurational entropy, driven by the resolution of energy scales corresponding to orders of degenerate perturbation theory on the approximately fragmented Hilbert space; it is only when these are averaged that a smooth growth ensues. Based on the statistics of jump times, we conclude that the available numerics data for the entropy growth, including our largest-scale MPS simulations with $N=50$ spins, cannot discriminate between a power-law or a single-logarithm time dependence.  This underscores the difficulty of identifying sub-logarithmic growth based on purely numerical data.
\zp{Effective models based on random unitary circuits, similar to Ref.~\cite{han2024exponentiallyslowthermalizationrobustness}, could further extend the timescales in the numerical simulations or provide analytical insights into the functional form of the entanglement growth.}
While double-logarithmic growth can arise in quenched-disorder MBL models, either as a subleading correction to entanglement entropy~\cite{Znidaric_MBL_2018} or as a leading-order term in particle number fluctuations in a subsystem~\cite{Emmanouilidis_doubl_log_PRB,Sieran2022,Ghosh2022,chavez2023ultraslowgrowthnumberentropy}, these effects appear unrelated to the ultraslow entanglement growth in the lattice Schwinger model. 

We note that the model~\eqref{eq:fermion_ham} has been realized with trapped ions~\cite{Martinez2016, Muschik2017, Kasper2017, Kokail2019}, while the 
configurational and number entropies have been experimentally measured in  ultracold ${}^{87}\mathrm{Rb}$ atoms in an optical lattice~\cite{Lukin2019}. Thus, quantum simulation platforms could provide further insights into the relation between ergodicity breaking in disordered MBL and the lattice Schwinger model.
In particular, current quantum simulation platforms allow the implementation of the tunable topological $\theta$-term \zp{that we use here to realize the deconfined phase, and which is responsible for a host of other exotic phenomena such as dynamical topological phase transitions and Coleman's phase transition~\cite{Jad_PRX_theta_term, Surace_PRXQ_theta_gauge}}.
By tuning the $\theta$-term, distinct forms of ergodicity-breaking have been realized in $U(1)$ quantum link models~\cite{Desaules2024ergodicitybreaking, Surace_PRXQ_theta_gauge}.
It would be interesting to \zp{further} explore the effect of the $\theta$-term on the relation between Hilbert space fragmentation and the rate of entanglement growth in the Schwinger model.
Finally, while we have focused on a $U(1)$ LGT in this work, it would be interesting to explore the dynamical effects of fragmentation in \emph{non-Abelian} LGTs, where forms of \emph{weak} ergodicity breaking have recently been identified~\cite{Ebner2024,Calajo2024}. 
 
\section{Methods}

\subsection{\zp{The structure of the Hilbert space}}\label{sec:model}

After a Jordan-Wigner transformation~\cite{SupplementaryInformation}, Gauss' law allows us to sequentially integrate out the gauge fields, leading to a Hamiltonian dependent on the background charge sector. In the spin formulation, for each charge sector $\{q_\alpha\}$,  the Hamiltonian breaks into two charge-independent terms and a charge-dependent term, given in Eqs.~\eqref{eq:spin_ham_XY}-\eqref{eq:spin_ham_charge_background} of the main text.  Formally, this corresponds to an $XY$ spin model with a local $Z$-field and a long-range $ZZ$ coupling. The latter breaks integrability and makes the interactions spatially-asymmetric, with the right-most spin on the lattice completely decoupled from the interaction. The last term of the Hamiltonian, $\Hq$ in Eq.~\eqref{eq:spin_ham_charge_background}, explicitly depends on the distribution of background charges $\{q_{\alpha}\}$.
For the special case of all background charges $q_{\alpha}$ being zero, this term takes the form:
\begin{multline}
    \zp{\hat H_{q_\alpha{=}0} = \frac{J}{2} \sum_{j=1}^{N} \left(\frac{\theta}{\pi} (N-j) - \left\lceil \frac{N - j}{2} \right\rceil \right) \sz_j}\\ +\frac{m}{2}\sum_{j=1}^{N} (-1)^j \sz_j\ ,
\end{multline}
and therefore represents a form of tilted potential.
The non-zero background charges can then be viewed as adding disorder to these local fields.

The spin Hamiltonian in Eq.~\eqref{eq:spin_ham} retains the original symmetries of the LGT.
The Coulomb and disorder terms in the Hamiltonian are purely diagonal, whereas the purely off-diagonal term $\hat{H}_{\pm}$ creates and annihilates electron-positron pairs, thus conserving the total particle number.
In the spin language, the Hamiltonian conserves the total magnetization $\sum_{i = 1}^N \sz_i$ corresponding to a $U(1)$ symmetry.
This is expected as the original continuum theory, QED in $1+1$D has a $U(1)$ gauge symmetry.
We resolve this symmetry by working in the sector with zero magnetization, or equivalently the half-filling sector of fermions, which has a Hilbert space dimension of $\binom{N}{N/2}$. 
In this sector, the model can have a further symmetry corresponding to charge conjugation and spatial reflection~\cite{Kokail2019}.
However, because of the presence of background charges in $\Hq$, this is only a symmetry if the background charges $\{q_{\alpha}\}$ themselves are symmetric under charge conjugation and reflection.
For example, this is a symmetry when all background charges are set to zero.
For randomly picked charge distributions, however, this is almost never the case and can be ignored.

As mentioned in the main text, the full state of the system is assumed to be a tensor product of the fermionic vacuum state, 
\begin{equation}\label{eq:fermion_init_state}
\ket{\Psi}_f = \ket{101010\dots} \equiv  \ket{\text{vac}}_f,
\end{equation}
and the superposition of the electric-field eigenstates:
\begin{equation}\label{eq:gauge_init_state}
    \ket{\Psi}_g = \bigotimes_{n = 1}^{N-1}\ket{\Bar{L}_n},\quad \ket{\Bar{L}_n} = \frac{\ket{-1}_n + \ket{0}_n + \ket{+1}_n}{\sqrt{3}},
\end{equation}
where $\hat{L}_n \ket{a}_n = a \ket{a}_n , a \in \mathbb{Z}$.

The choice of the fermionic initial state dictates the relationship between the gauge fields $\hat{L}_n$ and the background charges $q_n$ through the relation:
\begin{equation}
    \hat{L}_n = \hat{L}_{n-1} + q_n + [\langle \sz_n \rangle_{\Psi_f} + (-1)^n]/2.
\end{equation}
The last term vanishes for the N\'eel state of fermions in Eq.~\eqref{eq:fermion_init_state}, leading to the relation $q_n = \hat{L}_{n} - \hat{L}_{n-1}$.
Thus, the state of the gauge fields $\ket{\Bar{L}_n}$ is fully specified in this case by its background charges, which we denote by $\ket{q_n}$.
The full initial state then takes the form
\begin{equation}
    \ket{\Psi}_0 = \frac{1}{\sqrt{\mathcal{N}}} \sum_{\{q_n\}} \ket{\text{vac}}_f \otimes \ket{q_n}.
\end{equation}
Since each charge sector admits a Hamiltonian $\hat{H}_{\{ q_n\}}$, the sectors are disconnected and time evolution of the full wave function is described by the average over all sectors:
\begin{equation}\label{eq:time_evolution}
    \ket{\Psi(t)} = \frac{1} {\sqrt{\mathcal{N}}}\sum_{\{q_n\}} e^{-i\hat{H}_{\{q_{n}\}}t} \ket{\text{vac}}_f \otimes \ket{q_n}.
\end{equation}
As pointed out in~\cite{brenes2017many}, the sum in Eq.~\eqref{eq:time_evolution} then acts as an effective disorder average.
For simplicity and to match previous work, rather than calculating $q_n$ from the fields $L_n$, we draw $q_n$ directly from $\left\{-1, 0, +1\right\}$ with uniform probability;
we then impose the condition that $\sum_n q_n = 0$, which corresponds to fixing ${\hat{L}_0 = \hat{L}_N}$ such that there is no net electric charge in the system.
Note that restricting to three values on each site leads to the number of background charge sectors scaling as $\order{3^{N-1}}$.
In principle, the gauge field eigenvalues are unbounded and we could draw the charges $q_n$ from a broader window.
This should have effects similar to increasing the strength of $\Hq$ alone, increasing the strength of fragmentation.
In the SI~\cite{SupplementaryInformation}, we show the effects of scaling $\hat{H}_{ZZ}$ and $\Hq$ independently.
On the other hand, if we reduced $J$ while broadening the range of $q$, we would approach something closer to the uniform disorder seen in typical models of MBL.

\subsection{Degenerate perturbation theory}\label{sec:dpt_methods}

To describe dynamics at intermediate times, we developed a degenerate perturbation theory (DPT) approach following Ref.~\cite{Michailidis2018}, applicable in the limit $J/w \to \infty$. In this limit, the diagonal part of the Hamiltonian dominates, and we can expand the effective Hamiltonian within a tower $\mathcal{K}_0$, in orders of $w/J$:
\begin{equation}
	\hat{H}_\text{eff} = \hat{H}^{[0]} + \hat{H}^{[1]} + \hat{H}^{[2]} + \dots \label{eq:dpt_effective_ham}
\end{equation}
We denote evolution under DPT up to $n$th order as DPT($n$).
The leading term $\hat{H}^{[0]}$ is simply the diagonal parts of the Hamiltonian, $\hat{H}_{ZZ} + \Hq$, projected into our tower $\mathcal{K}_0$: by definition, this is a constant.
$\hat{H}^{[1]}$ is then the projection of $H_\pm$ into $\mathcal{K}_0$, which gives us the Krylov subspaces.
Therefore, DPT(1) is simply the dynamics assuming perfect Hilbert space fragmentation.
The higher order terms $\hat{H}^{[2]}$ and $\hat{H}^{[3]}$ then provide diagonal and off-diagonal corrections respectively to $\hat{H}_\text{eff}$, with the latter in particular connecting different subspaces within $\mathcal{K}_0$.
The derivation of $\hat{H}_\text{eff}$ up to $\hat{H}^{[3]}$ in provided in the SI~\cite{SupplementaryInformation}.

We note that the full $\hat{H}_\text{eff}$ will generate dynamics equivalent to evolution in the full Hilbert space followed by projection into $\mathcal{K}_0$,
\begin{equation}
	\hat{P}_0 e^{-i\hat{H}t} \ket{\Psi} = e^{-i\hat{H}_\text{eff}t} \hat{P}_0 \ket{\Psi}\ , \label{eq:inf_order_DPT}
\end{equation}
which we refer to as ``infinite-order'' DPT, or $\mathrm{DPT}(\infty)$.
While this is much more computationally intensive, it is a useful reference point in helping us understand the dynamics of this system in the large-$J$ regime.

\begin{figure}[tbp]
	\includegraphics[width=\linewidth]{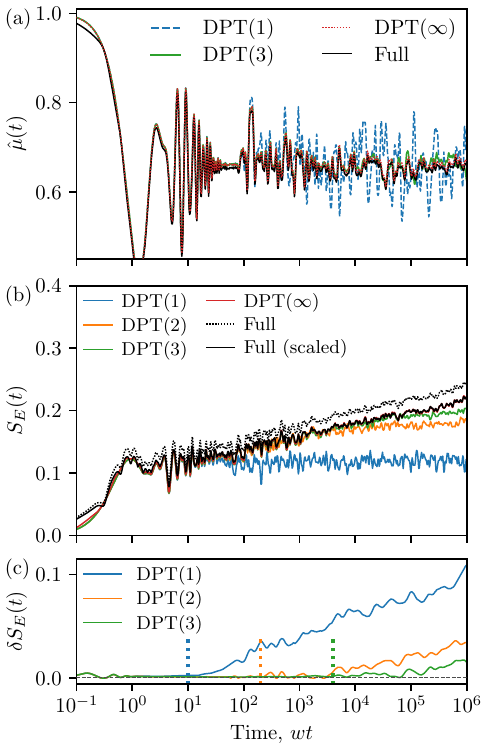}
	\caption{%
    (a) Staggered magnetisation $\hat{\mu}(t)$ following a quench from $\ket{\text{vac}}$, calculated using both 1st- and 3rd-order DPT .
    This is compared to ``infinite order'' DPT and full Hamiltonian evolution.
    We exclude 2nd-order DPT to avoid clutter.
    (b) Bipartite entanglement entropy $S_E(t)$, calculated using the first three orders of DPT, and again compared to DPT($\infty$) and full ED.
    We also show the full ED rescaled by a constant factor such that it agrees with DPT($\infty$) at $wt = 10^6$.
    (c) Error between finite-order DPT and DPT($\infty$), $\delta S_E(t)$.
    The vertical dotted lines indicate the rough times beyond which this error grows beyond a small threshold (grey dashed line); these times grow exponentially with the order of DPT.
    [In all cases, $N = 16$, $J / w = 10$, and we average over 256 charge sectors.]
    }
	\label{fig:dpt_entropy_growth}
\end{figure}
We will now compare DPT for the first three non-trivial orders, $n = 1, 2, 3$, to both DPT$(\infty)$ as well as full Hamiltonian evolution (``full ED'').
In Fig.~\ref{fig:dpt_entropy_growth}(a), we compute the staggered magnetization,
\begin{equation}
    \label{eq:stagged_magnetisation}
    \hat{\mu} = \frac{1}{N} \sum_{j} (-1)^j \sz_j\ ,
\end{equation}
following a quench from $\ket{\text{vac}}$.
In the ergodic phase, we expect $\hat{\mu}(t) \to 0$ as $t \to \infty$, but in the MBL phase this should decay to a finite value.
Even DPT(1) shows good agreement with both DPT($\infty$) and full ED for times up to \zp{$wt \approx 10^2$}, including capturing oscillations in this value at later times, although it fails to capture the \textit{decay} of these oscillations.
For this, we need to go to higher orders, and DPT(3) successfully describes this decay, \zp{agreeing with full ED up to $wt \approx 10^4$}.
We also see that DPT($\infty$) agrees well with full ED at all times, showing that the dynamics within a $J$-tower accurately describe the dynamics within the full Hilbert space for local observables.

In Fig.~\ref{fig:dpt_entropy_growth}(b), we then show a similar comparison for the bipartite entanglement entropy, $S_E(t)$ \eqref{eq:entanglement_entropy}.
This is a highly non-local observable, with a strong dependence on the fine structure of the quantum state, and should be much more challenging to capture.
Additionally, in Fig.~\ref{fig:dpt_entropy_growth}(c), we show the absolute difference between each order of DPT and DPT($\infty$), which we take as the error in the method and label $\delta S_E(t)$.
We see that each successive order of DPT is able to faithfully capture the entanglement entropy attained by DPT($\infty$) for a factor of $\mathcal{O}(2J/w)$ longer in time, up to about $wt \approx 5000$ for DPT(3).
We find that these scalings hold as we vary $J / w$.
However, full ED shows a marked difference from DPT($\infty$), showing that dynamics outside the initial $J$-tower contributes to the entanglement.
Despite this, if we rescale the full ED $S_E(t)$ data by a constant factor such that it agrees with DPT($\infty$) at $wt = 10^6$, then we observe almost perfect agreement between these two methods.
This suggests that, while dynamics outside the initial energy level are important, they are both qualitatively and quantitatively similar and do not change the overall functional form, meaning that DPT is able to fully describe the dynamics of the model at short to intermediate times.

We note also that the \zp{initial growth} in entanglement entropy at $wt < 1$ is captured even in $\text{DPT}(1)$.
This suggests that it is entirely due to dynamics in active regions which cross the bipartition; because hopping processes elsewhere in the chain do not affect the resonance condition Eq.~\eqref{eq:resonant_hop_condition}, these distant parts do not interact and so cannot generate entanglement at this order of perturbation theory.
Inspecting this more closely, only some charge sectors exhibit this \zp{growth}, each with an amplitude of $\ln2$ as would be expected due to the delocalization of a single particle across a partition.
The other charge sectors remain at (nearly) zero entanglement, which results in a much smaller spike after averaging over sectors.

\subsection{Matrix product state simulations}\label{sec:mps}

The $U(1)$ LGT Hamiltonian $\hat{H}$ in Eq.~\eqref{eq:spin_ham} can be written as a compact matrix product operator (MPO)~\cite{Schollwock2011,CiracRMP}.
Even though $\hat{H}_{ZZ}$ is long-ranged and non-uniform, it couples all sites with a strength that depends solely on the location of the right-hand site. Therefore, $\hat{H}$ can be expressed straightforwardly as an MPO:
\begin{equation}
	\hat{H} = \sum_{\{\alpha, \beta\}} \delta_{\alpha_1, 0} \left(\prod_{\ell = 1}^{N} A_{\alpha_\ell, \beta_\ell}^{[\ell]}\right) \delta_{\beta_N, 3}\ ,
\end{equation}
with $A^{[\ell]}$ a rank-$4$ tensor with entries
\begin{equation}
	A^{[\ell]} = \begin{pmatrix}
		\mathbb{I}		& w \spl	& w \smi	& (J/2) \sz		& h_k	\sz \\
		0				& 0			& 0			& 0				& \smi \\
		0				& 0			& 0			& 0				& \spl \\
		0				& 0			& 0			& \mathbb{I}	& (N - \ell) \sz \\
		0				& 0			& 0			& 0				& \mathbb{I}
	\end{pmatrix}\ .
\end{equation}
We calculate the numerical time evolution of the initial vacuum state using the time-dependent variational principle (TDVP) algorithm~\cite{haegeman2016,paeckel2019,mptoolkit}.
We use single-site updates with a dynamically growing bond dimension, fixing the density-matrix truncation threshold.
As the MPS bond dimension is continually increasing due to the build-up of entanglement, the computational effort required to perform a single timestep will increase as well, which limits the accessible evolution time for a fixed amount of computation time per charge sector.
Furthermore, since each charge sector will have a different entanglement growth rate, the charge sectors with the most rapid growth will limit the largest time for which we can calculate the average over all sectors.

\section{Data Availability}

Statement of compliance with EPSRC policy framework on research data: This publication is theoretical work that does not require supporting research data.

\section{Acknowledgments}	

We would like to thank Andrew Hallam and Jean-Yves Desaules for collaboration during the early stages of this project and for useful comments on the manuscript. 
We acknowledge support by the Leverhulme Trust Research Leadership Award RL-2019-015 and EPSRC grant EP/W026848/1. 
This research was supported in part by grant NSF PHY-2309135 to the Kavli Institute for Theoretical Physics (KITP).
J.C.H.~acknowledges support from the Max Planck Society.
Computational portions of this research work were carried out on ARC3 and ARC4, part of the High-Performance Computing facilities at the University of Leeds, UK.
Exact diagonalization simulations were performed using QuSpin 0.37~\cite{Quspin_1, *Quspin_2}.
Matrix product state simulations were performed on The University of Queensland's School of Mathematics and Physics Core Computing Facility \texttt{getafix}.

\section{Author contributions}

All authors contributed to the development of the ideas and analysis of the results.
J.J.\ and T.B.\ performed exact diagonalization calculations and designed the figures.
The MPS simulations were performed by J.J.O.\ and J.C.H.
J.J., T.B., and Z.P.\ wrote the manuscript with input from other authors.

\section{Competing interests}

The authors declare no competing interests.

\bibliography{biblio.bib}

\newpage 
\cleardoublepage

\startcontents[si]
\begin{center}
	\textbf{\large Supplemental Information}\\[5pt]
\end{center}

\setcounter{equation}{0}
\setcounter{figure}{0}
\setcounter{table}{0}
\setcounter{page}{1}
\setcounter{section}{0}
\renewcommand{\theequation}{S\arabic{equation}}
\renewcommand{\thefigure}{S\arabic{figure}}
\renewcommand{\thesection}{S\Roman{section}}
\renewcommand{\thepage}{\roman{page}}
\renewcommand{\thetable}{S\arabic{table}}

\printcontents[si]{}{1}[2]{} 
	

\section{Derivation of the lattice gauge theory}
\label{app:lgt_derivation}

In this section, we revisit the derivation of the lattice Schwinger model.
As a model of Quantum Electrodynamics (QED), the Schwinger model describes interactions between spinless fermions and antifermions via electric fields in one spatial dimension.
In this case, the vector potential $\hat{A}(x)$ has a temporal and a spatial component given by $(\hat{A}_0(x), \hat{A}_1(x))$.
We fix a gauge by setting $\hat{A}_0(x) = 0$.
In this gauge, the electric field operator becomes $\hat{E}(x)= -\partial_0\hat{A}_1(x)$ where $\partial_0$ is the partial time derivative.  The electric field operator is the canonical momentum conjugate to the vector potential $\hat{A}(x)$ with the commutation relation $[\hat{A}_1(x), \hat{E}(x')] = -i\delta(x - x')$.
The model consists of a matter field $\hat{\Phi}(x)$ which is a two-component spinor field $(\hat{\Phi}_{e^-}(x), \hat{\Phi}^{\dag}_{e^+}(x))^T$ representing electrons and positrons.

In the continuum limit with natural units $\hbar = c = 1$, the Schwinger model has a $U(1)$ gauge symmetry and the Hamiltonian is given by:
\begin{multline}\label{eq:continuum_ham}
	\hat{H}_{\text{cont}} = \int dx \biggl( -i \bar{\Phi}(x) \gamma^1 D_{1} \hat{\Phi}(x)\\
	+ m \bar{\Phi}(x) \hat{\Phi}(x) + \frac{1}{2} \hat{E}^2(x) \biggr)\ ,
\end{multline}
where $D_{1} = \partial_1 - ig\hat{A}_1(x)$ is the gauge covariant derivative in one spatial dimension, $\partial_1$ is the partial derivative with respect to $x$, and $m$ is the fermionic mass.
The Dirac adjoint of the spinor field  is defined as $\Bar{\Phi} = \hat{\Phi}^{\dag}\gamma^0$, where $(\gamma^0,\gamma^1) = (\sz, i\sy)$ are the Dirac matrices in 1+1-dimension.
We  set the coupling constant $g = e$ as the charge $e$ of the electrons.

Moving away from the continuum, this model can also be formulated on a one dimensional lattice where points in space are separated by a distance $a$ and time is continuous.
In particular, we use the lattice version first developed by Kogut and Susskind~\cite{Kogut_Susskind_LGT_PRD_1975, Banks_Sussking_Sch_PRD_1976}. In this formulation, the two-component matter field $ (\hat{\Psi}_{e^-}(x), \hat{\Psi}^{\dag}_{e^+}(x))^T $ is ‘unfolded’ onto the lattice, by placing the electron/positron fields onto alternating even/odd sites, with two neighboring sites defining a unit cell.
The discrete versions of the vector potential and the electric fields are placed on the links connecting neighbouring lattice sites.

To respect the $U(1)$ symmetry of the original model, we move from the canonically conjugate continuous fields $\hat{E}(x)$, $\hat{A}_1(x)$ to their corresponding discrete versions $\hat{\phi}_n$, $\hat{L}_n$ such that the canonical commutation relations are
\begin{equation}
	\label{eq:commutation_gauge_field}
	\comm*{\hat{A}_1(x)}{\hat{E}(x')} = -i \delta(x - x'), \quad \comm*{\hat{\phi}_n}{\hat{L}_m} = i \delta_{n, m},    
\end{equation}
where the $\hat{\phi}_n = a g \hat{A}_1(x_n)$ is a $U(1)$ parallel transporter and $\hat{L}_n = (-1/g)\hat{E}(x_n)$ is the canonically conjugate electric field defined on the link between neighboring sites $(n, n+1)$.
The vector potential $\hat{A}_1(x_n)$ enters the Hamiltonian through an exponential term of the form $e^{i \hat{\phi}_n} = e^{-i a g \hat{A}_1(x_n)}$, which requires $0 \leq \phi(n) \leq 2 \pi$, where 
$\phi(n)$ represents the eigenvalue of $\hat{\phi}_n$.
Looking at the commutation relation between $\hat{\phi}_n$ and $\hat{L}_n$, it is straightforward to see that $\hat{L}_n$ generates cyclic translations in $\hat{\phi}_n$ and is thus an angular momentum operator.
As a consequence of the range of $\hat{\phi}_n$ and Eq.~\eqref{eq:commutation_gauge_field}, $\hat{L}_m$ is quantized with integer eigenvalues,
\begin{equation}
	\hat{L}\ket{L} = \hat{L} \ket{L},\quad \hat{L} = 0, \pm1, \pm 2...
\end{equation}
Then, $e^{i \hat{\phi}_n}$ acts as a ladder operator on the eigenstates $\ket{l}$ of $\hat{L}_m$ as:
\begin{equation}
	e^{\pm i \hat{\phi}} \ket{l} = \pm \ket{l \pm 1}.
\end{equation}
We define one component fermionic field operators on each site as
\begin{equation}
	\newlength{\foreven}
	\settowidth{\foreven}{\text{for even}}
	\hat{\Psi}_n =
	\begin{cases}
		\sqrt{a} \hat{\Phi}_{e^-}(x_n) & \text{for even}~n\\
		\sqrt{a} \hat{\Phi}^{\dag}_{e^+}(x_n) & \makebox[\foreven][l]{\text{for odd}}~n\ .
	\end{cases}
\end{equation}
We can now write down the lattice Schwinger Hamiltonian on a chain of $N$ sites with open boundary conditions by replacing the integrals in Eq.~\eqref{eq:continuum_ham} with discrete sums as
\begin{equation*}
	\begin{aligned}
		\hat{H}_\text{sch} &= -iw \sum_{n=1}^{N-1}[\hat{\Psi}_n^{\dag} \hat{U}_n \hat{\Psi}_{n+1} - \text{H.c.}] \\
		&+ J\sum_{n=1}^{N-1} \left(\hat{L}_n + \frac{\theta}{2\pi}\right)^2
		+m\sum_{n=1}^N(-1)^n\hat{\Psi}_n^{\dag} \hat{\Psi}_n\ ,
	\end{aligned}
	\tag{\ref{eq:fermion_ham} revisited} 
\end{equation*}
with $\hat{U}_n = e^{i \hat{\phi}_n}$.
The first term in \eqref{eq:fermion_ham} describes nearest-neighbor hopping and corresponds to the creation and annihilation of electron-positron pairs.
The strength of this interaction is given by $w = \nicefrac{1}{2a}$.
The second term describes the rest mass energy (proportional to the fermionic mass $m$), while the last term describes the energy stored in the electric fields with $J = g^2 a/2$.
The lattice Hamiltonian reduces to the original Schwinger Hamiltonian in Eq.~\eqref{eq:continuum_ham} in the limit ${a \rightarrow 0}$.
We have further included a constant background field $\theta / 2 \pi$, also known as the topological $\theta$ term.

States in the Hilbert space are constrained by Gauss' law, which takes the form $\partial_1 \hat{E}(x) = g \Bar{\Phi}\gamma^0\hat{\Phi}$ in the continuum limit.
In the lattice version, Gauss' law can be formulated by considering a set of on-site generators
\begin{equation}
	\hat{G}_n = \hat{L}_{n} - \hat{L}_{n-1} - \hat{\Psi}_n^{\dag}\hat{\Psi}_n + \frac{1}{2}[1 - (-1)^n]    
\end{equation}
such that $[\hat{H}_\text{sch}, \hat{G}_n] = 0$.
\textit{Physical} states in the Hilbert space are defined as eigenvectors of $\hat{G}_n$, such that,
\begin{equation}
	\hat{G}_n \ket{\psi_\text{phys.}} = q_n \ket{\psi_\text{phys.}}\ ,
\end{equation}
where ${q_n}$ represents the distribution of background charges.
The fact that $\hat{G}_n$ commutes with the Hamiltonian has an important consequence on the dynamics of the system: $\hat{H}_\text{sch}$ does not mix states with different $q_n$ during dynamics, thus $\hat{H}_\text{sch}$ is block-diagonal is the eigenbasis of $\hat{G}_n$.
The $q_n$'s represent static background charges in the theory that act as sources of electric flux.

\section{Jordan-Wigner mapping} \label{app:jordan_wigner}
We may recast the lattice Schwinger Hamiltonian from Eq.~\eqref{eq:fermion_ham} as a spin model by a Jordan-Wigner transformation of the fermionic fields.
Writing $\hat{\Phi}_n = \prod_{l < n}[i \sz_l] \smi_n$ ($\hat{\Phi}_n^\dag = \prod_{l < n}[-i \sz_l] \spl_n$) and using a gauge transformation $\spm_n \Rightarrow \prod_{l<n}[e^{\mp i \hat{\phi}_l}] \spm_n$,  we arrive at the spin Hamiltonian
\begin{equation}\label{eq:spinham}
	\begin{split}
		\hat{H}_\text{spin} &= \hat{H}_\pm  
		+ \frac{m}{2} \sum_{n = 1}^N (-1)^n \sz_n 
		+ J\sum_{n=1}^{N-1} \hat{L}_n^2, \\
		\hat{H}_\pm &= w \sum_{n = 1}^{N-1} ( \spl_n \smi_{n+1}  + \text{H.c.}). 
	\end{split}
\end{equation}
Now, we integrate out the gauge fields using Gauss' law.
In the spin language, the Gauss' law constraint becomes $\hat{L}_n - \hat{L}_{n-1} = \frac{1}{2}[\sigma_n^z + (-1)^n] + q_n$.
By setting the boundary fields $\hat{L}_0 = \hat{L}_{N} = 0$, we get
\begin{multline}
	\sum_{n = 1}^{N-1} \left(\hat{L}_n + \frac{\theta}{2\pi}\right)^2\\
	= \sum_{n = 1}^{N-1}\Bigg(\sum_{m = 1}^n \Big[ \frac{1}{2}[\sz_m + (-1)^m] + q_m\Big] + \frac{\theta}{2\pi} \Bigg)^2\ , 
\end{multline}
where we have now included the topological background field $\theta$.
Representing the energy stored in the electric fields, this term can be separated into two parts consisting of a long-ranged two-body term $\hat{H}_{ZZ}$, and a field term $\Hq$ dependent on the charge configuration:
\begin{equation}
	\begin{split}
		\hat{H}_{ZZ} &= \frac{J}{2} \sum_{j=1}^{N-2} \sum_{k = j+1}^{N - 1}(N - k) \sz_j \sz_k\ , \\
		\Hq &= \frac{m}{2} \sum_{j=1}^{N} (-1)^j \sz_j \\
		&+ \frac{J}{2} \sum_{k = 1}^{N - 1} \sum_{j = k}^{N - 1} \left[2 \left(\sum_{i = 1}^{j} q_i\!\right)\! - (j \!\!\! \mod{2}) + \frac{\theta}{\pi}\right] \sz_k\ .
	\end{split}
\end{equation}
Curiously, the long-range term $\hat{H}_{ZZ}$ which represents Coulomb interactions and the term contributing to onsite disorder $\Hq$ both originate from the same term in the original Hamiltonian.
Unlike the conventionally studied MBL models such as the random field XXZ chain where interactions and disorder can be tuned independently, a single parameter $J$ controls both the interactions and the strength of disorder in this model.

\section{Derivation of the resonance condition}\label{app:resonance_derivation}

Let us examine the energy shift when we interchange two spins at sites $\ell$ and $\ell + 1$, $\ket{01} \to \ket{10}$, under the action of the $\hat{H}_\pm$ term in Eq.~\eqref{eq:spinham}.
First, noting that $\sz_\ell \sz_{\ell + 1}$ stays constant, let us explicitly write out all terms in $\hat{H}_{ZZ}$ which contribute to the shift:
\begin{multline}
	\Delta\hat{H}_{ZZ} = \frac{J}{2}\left(\sum_{j=1}^{\ell - 1} (N - \ell) \sz_j \sz_\ell + \sum_{j=1}^{\ell - 1} (N \! - \ell \! - 1) \sz_j \sz_{\ell + 1}\right.\\
	\left. + \sum_{k=\ell + 2}^{N - 1} (N-k) \sz_{\ell} \sz_k + \sum_{k=\ell + 2}^{N - 1} (N-k) \sz_{\ell + 1} \sz_k \right)\ .
\end{multline}
We can see that the second line is invariant under $\sz_{\ell} \leftrightarrow \sz_{\ell + 1}$, and so we can reduce to the first line.
Factorizing this, we obtain
\begin{equation}
	\begin{aligned}
		\Delta \hat{H}_{ZZ} &= \frac{J}{2} \left[(N \! - \! \ell) \sz_{\ell} + (N \! - \! \ell \! - \! 1) \sz_{\ell + 1} \right] \sum_{j=1}^{\ell - 1} \sz_j\\
		&= \frac{J}{2} \sz_\ell \, \sum_{j=1}^{\ell - 1} \sz_j\ ,
	\end{aligned}
\end{equation}
where we have used the fact that $\sz_{\ell} + \sz_{\ell + 1} = 0$.
Since $\sz_\ell$ goes from $-1$ to $+1$, the energy change of this hop due to $\hat{H}_{ZZ}$ is then $J \sum_{j=1}^{\ell - 1} \sz_j$.
At the same time, $\Hq$ changes by $2(m + h_\ell - h_{\ell + 1})$.
Therefore, substituting in the expression for $h_k$, we obtain the condition for resonant hopping with $m = 0$,
\begin{equation*}
	\frac{\Delta E}{J} \equiv \sum_{j=1}^{\ell - 1} \sz_j + 2 \sum_{j=1}^\ell q_j + \frac{\theta}{\pi} - (\ell \!\!\! \mod 2) = 0\ .
	\tag{\ref{eq:resonant_hop_condition} revisited}
\end{equation*}

We may also see that, when the resonance condition is not met and $\theta / \pi = \pm 1$, the energy always changes in increments of $2J$.
This follows since $J \sum_{j=1}^{\ell - 1} \sz_j - (\ell \!\!\! \mod 2)$ is always odd, while $2 \sum_{j=1}^\ell q_j$ is evidently even, and so the LHS of Eq.~\eqref{eq:resonant_hop_condition} is even if $\theta / \pi$ is odd.
For all other values of $\theta$, the resonance condition can never be satisfied, and in particular for $\theta / \pi = 0$, $\Delta E$ is at least $J$.

\section{Degenerate perturbation theory for the lattice Schwinger model}\label{app:dpt_derivation}
In the main text we have introduced the degenerate towers $\left\{\mathcal{K}_a\right\}$, with $\mathcal{K}_0$ denoting the tower containing our initial state, each with an unperturbed energy $E_a$.
We can also define $\left\{\mathcal{K}_a^b\right\}$ as the set of Krylov subspaces within $\mathcal{K}_a$, which are generated by $\hat{H}^\pm$ projected into $\mathcal{K}_a$, and likewise let $\mathcal{K}_0^0$ contain our initial state.
Let $\hat{P}_a$ and $\hat{P}_a^b$ be the projectors onto these towers and subspaces, respectively.
We then define off-diagonal blocks,
\begin{equation}
	\T{i}{j} = \hat{P}_j^{\mathstrut} \hat{H}_{\pm} \hat{P}_i^\dag\ .
\end{equation}
Note that $\Tdag{i}{j} = \T{j}{i}$.
We can then write down,
\begin{gather}
	\hat{H}^{[0]} = \hat{P}_0 \left(\hat{H}_{ZZ} + \Hq\right) \hat{P}_0^\dag\ ,\\
	\hat{H}^{[1]} =  \hat{P}_0^{\mathstrut} \hat{H}_{\pm} \hat{P}_0^\dag = \T{0}{0}\ ,
\end{gather}
representing the full Hamiltonian projected onto the tower $\mathcal{K}_0$.
Higher-order corrections may then be calculated through successive Schrieffer-Wolff transformations~\cite{Bravyi2011}, giving the effects of virtual hops to other energy levels.
The second- and third-order contributions are given by:
\begin{gather}
	\hat{H}^{[2]} = \sum_{a \neq 0} \frac{\T{0}{a} \T{a}{0}}{E_a - E_0}\ ,\\
	\begin{split}
		\hat{H}^{[3]} = &\sum_{a \neq 0} \frac{2 \T{0}{a} \T{a}{a} \T{a}{0}- (\T{0}{0} \T{0}{a} \T{a}{0} + \text{H.c.})}{2 (E_a - E_0)^2}\\
		+ &\sum_{a \neq b} \Delta_{ab}^0 \T{0}{b}\T{b}{a}\T{a}{0}\ ,
	\end{split}\\
	\Delta_{ab}^0 = \frac{(E_b - E_a) + (E_b - E_0)}{(E_a - E_0)(E_b - E_0)(E_b - E_a)} + (a \leftrightarrow b)\ .
\end{gather}
$\hat{H}^{[2]}$ tends to be diagonal, generating dephasing between states within a Krylov subspace, while $\hat{H}^{[3]}$ and higher generally also act off-diagonally to connect different Krylov subspaces at the same energy.

\section{Many-body Thouless parameter}\label{app:thouless}

The change in the nature of entanglement as a function of coupling ratio $J/w$ in the Schwinger model also has consequences for the response of eigenstates to local perturbations.
In the ergodic phase, a local perturbation strongly hybridizes an extensive number of eigenstates.
It was argued in Ref.~\cite{Serbyn15} that such perturbations should only hybridize an intensive number of eigenstates in an MBL phase, as a local perturbation can only affect the degrees of freedom within a fixed localization radius of its support.
For a local perturbation $\hat{V}$, the extent of the such hybridization can be probed using a dimensionless parameter -- a many-body generalization of the Thouless conductance:
\begin{equation}
	\mathcal{G}(\epsilon, N) = \ln \frac{|V_{n, n'}|}{E_{n+1}' - E_n'},
\end{equation}
where $\epsilon = E_n'/N$ are the sorted energy densities of the perturbed Hamiltonian, and $V_{n, n'}$ are the matrix elements of the perturbation operator in the unperturbed eigenbasis.

\begin{figure}[tbp]
	\includegraphics[width=1\linewidth]{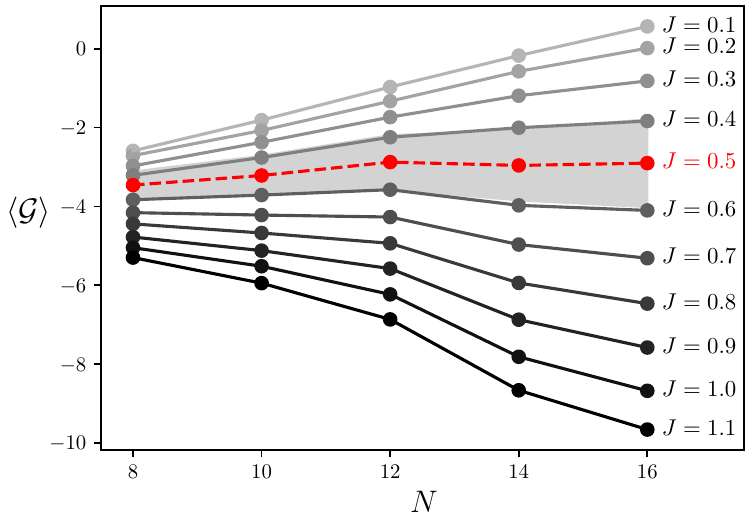}
	\caption{
		The Thouless parameter $\mathcal{G}$, averaged over charge sectors, as a function of system size for different values of $J$ and fixed $w=1$.
		The averaging is done across 500 charge sectors for $N<16$ and 450 charge sectors for $N = 16$.
		The perturbation is a local $\sz$ operator on the first site with strength $w/10$.
		The Thouless parameter is seen to become independent of system size around $J \approx 0.5$ (red dashed line). The shaded region represents the uncertainty (not to scale) in $J$ around the transition point. 
	}
	\label{fig:thouless_parameter}
\end{figure}

In an ergodic phase, a local operator strongly mixes neighboring eigenstates, leading to $\mathcal{G} \sim \mathcal{O}(N)$.
In the MBL phase, eigenstates with neighboring energies are coupled exponentially weakly, leading to $\mathcal{G} \sim -\mathcal{O}(N)$.
Then, the ergodic-MBL transition point can be identified as the $J$ where $\mathcal{G}(N) \sim \mathcal{O}(1)$.
In Fig.~\ref{fig:thouless_parameter}, we plot the averaged Thouless parameter as a function of system size for different values of $J$.
The averaging is over $33 \%$ of eigenstates around the DOS peak, after which we further average over charge sectors.
The results are shown for a local perturbation $\sz$ on the first site of the chain with a strength $w/10$, although the results are similar for different sites and using other perturbations such as $\hat{\sigma}_i^+ \hat{\sigma}_{i+1}^-$.
For small $J$, we observe an approximately linear increase of the Thouless parameter, consistent with an ergodic phase.
Upon increasing $J$, the dependence on system size changes sign, with a decay becoming visible for $J > 0.6$, consistent with an MBL regime.
We identify the region around $J \approx 0.5$ as the regime where the Thouless parameter becomes independent of system size, i.e., the ergodic-MBL transition. 

\section{Spectral form factor}\label{app:sff}

\begin{figure}[tbp]
	\includegraphics[width=\linewidth]{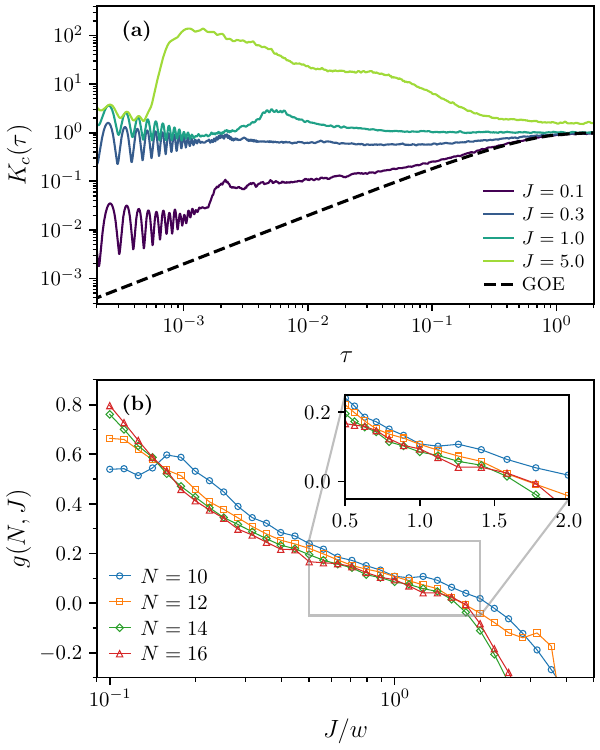}
	\caption{
		(a) The connected spectral form factor (SFF), $K_{c}(\tau)$, for selected $J$ at $N = 16$ (and $w = 1$).
		We also show the GOE ensemble value as the black dashed line.\linebreak[2]
		(b) $g(N, J) = \log_{10}(\tau_\text{H} / \tau_\text{GOE})$, which measures when the SFF converges to the GOE value, for various $N$ as a function of $J / w$.
		In the thermodynamic limit, we expect $g \to +\infty$ ($g \to -\infty$) for chaotic (non-ergodic) systems.
	}
	\label{fig:SFF}
\end{figure}
Eigenstates carry information about the time $t\rightarrow\infty$ limit, which is not directly accessible to numerics otherwise.
To probe the behavior of the system at finite times, we consider the Spectral Form Factor (SFF)~\cite{haake1991quantum,gutzwiller2013chaos}, defined as the Fourier transform of a two-point spectral correlation function.
Crucially, the SFF quantifies \textit{long-ranged} correlations between energy eigenvalues and thus holds more information than, e.g., the level spacing ratio $r$, which is only dependent on short-ranged spectral correlations.
Calculating the SFF requires ``unfolding'' the spectrum to make the density of eigenvalues equal to unity~\cite{Suntajs2020}, after which it is defined as
\begin{equation}
	K(\tau) = \frac{1}{Z} \left\langle \Biggl|  \sum_{j = 1}^{\mathcal{N}} e^{-2\pi i\epsilon_j \tau} \Biggr|^2 \right\rangle\ ,
\end{equation}
where $\{ \epsilon_1,..., \epsilon_D\}$ are the unfolded eigenvalues, $\tau$ is the rescaled time following the unfolding,
and $Z$ is a normalization factor chosen such that $K(\tau) \to 1$ as ${\tau \to \infty}$.
In these units, the Heisenberg time, which is given by the inverse of the mean level spacing, is $\tau_\text{H} = 1$.
$\tau_\text{H}$ measures when discrete energy levels can be resolved, after which the finite nature of the system becomes evident.
We furthermore choose to calculate the connected SFF, $K_c(\tau)$, which additionally subtracts a non-universal disconnected part,
\begin{equation}
	K_c(\tau) = K(\tau) - A \Biggl| \left\langle \sum_{j = 1}^{\mathcal{N}} e^{-2\pi i\epsilon_j \tau} \right\rangle \Biggr|^2\ ,
\end{equation}
with $A$ again a normalization factor chosen to ensure $K_c(\tau) \to 0$ as $\tau \to 0$.
This avoids the need to apply a Gaussian filter to the unfolded spectrum~\cite{Suntajs2020}.

The SFF, in principle, provides detailed information about the distribution of eigenvalues in the system; however, for our purposes, we are mostly interested in determining whether the system exhibits chaotic dynamics.
This can be done by comparing $K_c(\tau)$ to the GOE prediction, given by $K_{c,\,\text{GOE}}(\tau) = \tau (2 - \ln(1 + 2\tau))$ for $\tau < 1$.
In chaotic systems, $K_c(\tau)$ will agree with the GOE prediction after the (rescaled) Thouless time $\tau_\text{Th}$, which can be interepreted as the time after which the dynamics are universal and the system is indistinguishable from a random matrix.

In Fig.~\ref{fig:SFF}(a) we show the connected SFF for several values of $J$ and $N = 16$, alongside the GOE prediction.
For small but finite $J = 0.1$, the SFF rapidly approaches the random matrix form, and therefore we can classify the system as chaotic in this regime.
However, as we increase $J$, we see that the SFF deviates from the GOE prediction, and by $J = 2$ no longer intercepts it before $\tau_\text{H} = 1$, but instead approaches the constant value $K_c(\tau) = 1$ well before $\tau_\text{H}$; this is typical of the MBL regime~\cite{Suntajs2020}.
The SFF takes on a particularly unusual form for $J = 5$, with plateaus observable at $\tau = (2J/w)^{-2}$ and $(2J/w)^{-3}$: these likely follow from the spectral splitting observed in the main text, and is further evidence that the system may not follow the conventional MBL phenomenology in the strong-coupling regime. While Ref.~\cite{Giudici_MBL_U1} has already studied the SFF of the Schwinger model, the anomalous SFF features for large $J$ have not been pointed out.

To further quantify the crossover from chaotic to MBL dynamics, we calculate $g(N, J) = \log_{10}(\tau_\text{H} / \tau_\text{GOE})$ in Fig.~\ref{fig:SFF}(b), where $\tau_\text{GOE}$ is the last time at which $K_c(\tau)$ deviates significantly from the GOE prediction, i.e.,\ $|\log_{10}(K_c(\tau)/K_{c,\,\text{GOE}})| < 0.08$ for all $\tau > \tau_\text{GOE}$.
For chaotic systems, is expected that $g \to +\infty$ in the thermodynamic limit, indicating a rapid approach to random matrix dynamics; on the other hand, $g \to -\infty$ for non-ergodic systems, indicating that they do not obey the GOE at any time before finite size effects dominate.
However, in practice, $\tau_\text{GOE} > \tau_{H}$ ($g < 0$) is not physically meaningful, as beyond the Heisenberg time the discrete nature of the system is resolved and the dynamics cannot be related to that of the thermodynamic limit.
The procedure for computing $\tau_\text{GOE}$ leads to an unknown constant offset in $g(N, J)$, but we can identify the crossover as the region in which $g \to 0^+$, which here is $J_c \approx 1$: this is shown more closely in the inset of Fig.~\ref{fig:SFF}(b).

\section{Other initial states}\label{app:otherstates}

\begin{figure}[tbp]
	\includegraphics[width=0.9\linewidth]{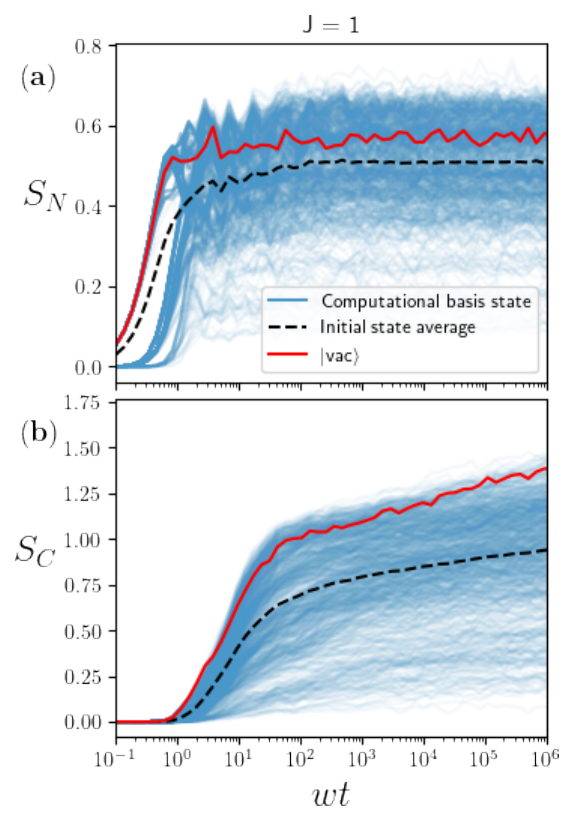}
	\caption{
		Growth of (a) number and (b) configurational entropy from all initial computational basis states for $J=1$ and $N = 12$.
		The entropy series for each initial state is averaged over 150 charge sectors.        
	}
	\label{fig:app_all_initial_states}
\end{figure}

In the main text, when we refer to thermalization or ergodicity breaking, we have in mind the generic behavior of the system from typical initial states. However, both our main text and previous literature \cite{brenes2017many} have focused on the dynamical properties of the Schwinger model prepared in a single $\ket{\text{vac}}$ initial state. How typical is the $\ket{\text{vac}}$ initial state?
Although $\ket{\text{vac}}$ is the bare vacuum of staggered fermions, it is not the ground state of the spin model, with its energy given by
\begin{equation}
	E_\text{vac} = -\frac{J}{2} \left(\frac{N^2}{4} - \sum_{k = 1}^{N - 1} \left\lceil \frac{N - k}{2} \right\rceil q_k \right)\ .
\end{equation}
For generic background charges $q_k$'s, we find that $E_{\text{vac}}$ is close to but smaller than the energy at the DOS peak and so we expect its dynamical behavior to be representative of a typical initial state.
In Fig.~\ref{fig:app_all_initial_states}(a) and (b), we plot the growth of the charge sector averaged entanglement entropy $S_E$ from all initial product states in the computational basis for $J/w = 1$, wherein we average over 150 charge sectors for each initial state.
Although most initial states show a quick growth and saturation in number entropy, we observe qualitatively different behavior in the growth of configurational entropy.
We find that $\sim 70\%$ of initial states show a relatively quick growth and saturation to steady-state values ($S_C > 0.9)$, whereas the remaining initial states show a slower growth of entanglement over large timescales.
The $\ket{\text{vac}}$ state behaves typically and lies in the former group of states with a faster growth of $S_C$.
Although the variability between initial states increases upon increasing $J/w$, the growth of entanglement from the $\ket{\text{vac}}$ state remains qualitatively similar to the majority of initial states. We note that at large-$J$, fragmentation occurs at the level of the Hilbert space, thus affecting the dynamics of all initial states. 

\section{Coulomb interactions versus disorder}
\begin{figure}[tbp]
	\includegraphics[width=\linewidth]{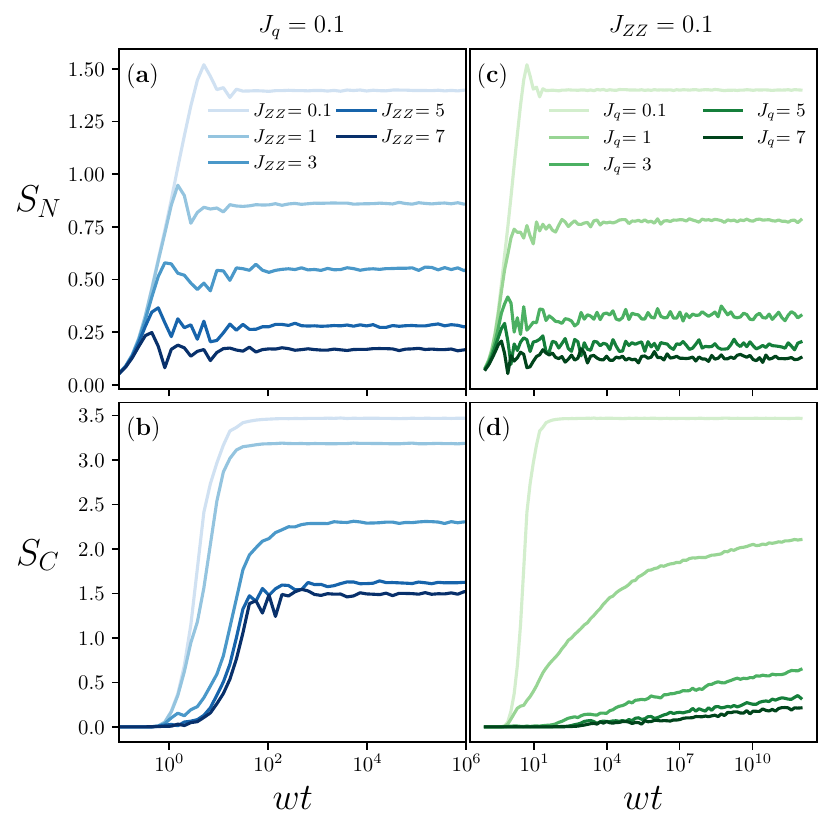}
	\caption{
		(a)-(b): Growth of the number and configurational entropy with a fixed disorder strength $J_q = 0.1$ and increasing Coulomb interaction strength $J_{ZZ}$.
		The entropy growth is logarithmic when disorder is small and Coulomb interactions dominate.
		(c)-(d): Growth of number and configurational entropy with a fixed Coulomb interaction strength $J_{ZZ} = 0.1$ and increasing disorder strength $J_q$.
		Ultraslow growth of the configurational entropy is seen when Coulomb interactions are small and disorder dominates.
		All data is for the $\ket{\text{vac}}$ initial state with $N = 16$, averaged over 25 charge sectors.
	}
	\label{fig:app_separate_magnitudes}
\end{figure}
In this section, we analyse the impact of the Coulomb interactions and disorder on the growth of entropy by tuning their magnitudes independently in the Hamiltonian. We write the full Hamiltonian as $\hat{H}_{\left\{q_\alpha\right\}} = \hat{H}_{\pm} + J_{ZZ}\hat{H}_{ZZ} + J_{q}\Hq$, where $\hat{H}_{ZZ}$ and $\hat{H}_{q}$ are the Coulomb and disorder terms with strengths $J_{ZZ}$ and $J_{q}$ respectively.
In Fig.~\ref{fig:app_separate_magnitudes}(a)-(b), we plot the number and configurational entropies starting from the $\ket{\text{vac}}$ initial state by first fixing a small disorder strength $J_q = 0.1$, and gradually increasing the Coulomb interaction strength $J_{ZZ}$.
Both the number and configurational entropies grow roughly logarithmically, followed by a saturation value that decreases with increasing Coulomb interaction strength.
Although the growth of the configurational entropy is slower, it still saturates by $wt \approx 10^2$ even for $J_{ZZ} = 10$.
This implies that although the growth and the corresponding saturation value of entanglement entropy is suppressed upon increasing the Coulomb interaction strength, the Coulomb interactions are not themselves responsible for the ultraslow growth of entropy observed at large $J$. 

The origin of the slow growth of entanglement entropy is instead attributed to the disorder term which is controlled by $J_q$. In Fig.~\ref{fig:app_separate_magnitudes}(c)-(d), we test this by setting the Coulomb interaction strength to a small value $J_{ZZ} = 0.1$ and gradually increasing the disorder strength $J_q$.
Upon increasing $J_q$, the growth of number entropy is strongly suppressed, although it still saturates at early times. The slow growth of configurational entropy starts to emerge at around $J_q = 3$, similar to that observed in the main-text. We note that both the terms $J_q$ and $J_{ZZ}$ induce ergodicity-breaking individually, and the level spacing ratio varies smoothly from Wigner-Dyson to Poisson and dips further below the Poisson value as the strength of either term is increased whilst keeping the other fixed. However the nature of ergodicity-breaking induced by both terms is distinct, as only the disordered term leads to an ultraslow growth of entanglement. The role of the discrete disorder term $J_q$ is studied in more detail in the next section.

\section{Comparison with disordered XXZ model}\label{app:xxz_comparison}
\begin{figure*}[tbp]
	\includegraphics[width=0.8\linewidth]{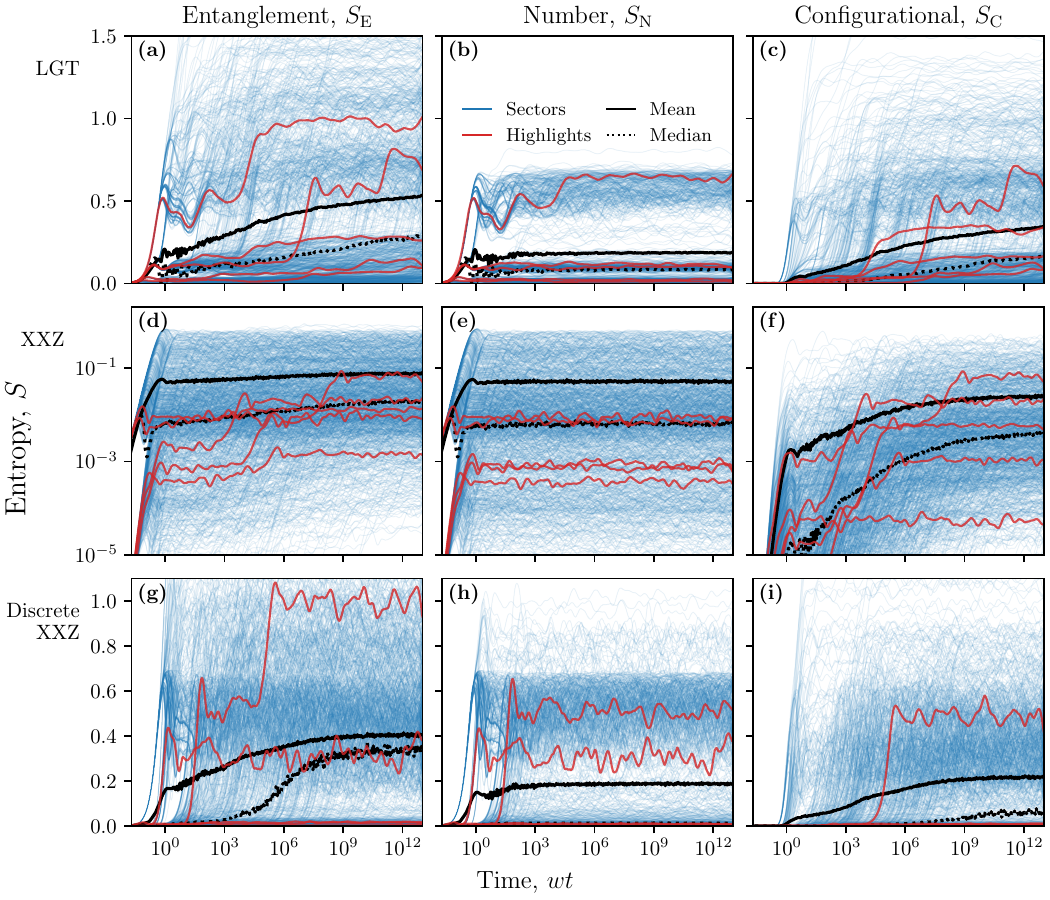}
	\caption{
		(a-c) The entanglement entropy $S_E$, number entropy $S_N$, and configurational entropy $S_C$, respectively, for the lattice Schwinger model with $J = 5$, following a quench from $\ket{\text{vac}}$.
		Results for 1000 charge sectors are shown in blue;
		five randomly-chosen trajectories have been highlighted in red to help guide the eye.
		We also show the charge sector mean (black solid line) and median (black dashed line).
		(d-f) The same, but for the uniform-disorder XXZ model \eqref{eq:XXZ_hamiltonian} with $W = 20$.
		(g-i) The same, but for the discrete-disorder XXZ model with $W_j \in \{-W, 0, +W\}$ and $W = 20$.
	}
	\label{fig:app_xxz_w20}
\end{figure*}
In Fig.~\ref{fig:app_xxz_w20}, we compare the behavior of the lattice Schwinger  model with that of the disordered XXZ model, the prototypical model of MBL:
\begin{equation}
	\label{eq:XXZ_hamiltonian}
	\begin{split}
		\hat{H}_\mathrm{XXZ} =& \sum_{j = 1}^{N - 1} J_{XY} \left(\spl_j \smi_{j + 1} + \smi_j \spl_{j + 1}\right)\\
		&+ \sum_{j = 1}^{N - 1}\frac{J_Z}{2} \sz_j \sz_{j + 1} + \sum_{j = 1}^{N} W_j \sz_j\ , 
	\end{split}
\end{equation}
where $W_j$ is the quenched disorder potential drawn uniformly from the interval $\left[-W, W\right]$.
Panels (a), (b) and (c) of Fig.~\ref{fig:app_xxz_w20} show the entanglement entropy $S_E$, number entropy $S_N$, and configurational entropy $S_C$,  respectively, for the Schwinger model with $J = 5$. We observe the characteristic slow growth in entanglement entropy, dominated at late times by the configurational entropy. We also observe the ``jump'' behavior in the configurational entropy, and banding in all three plots.

These can be compared to panels Fig.~\ref{fig:app_xxz_w20}(e-f), which depict the same for the XXZ model with $J_{XY} = J_Z = 1$ and $W = 20$.
Here, we see that the entanglement entropy very quickly saturates, and that the number entropy is dominant at all times.
The configurational entropy $S_C$ appears to exhibit some sort of slow growth, with a much smaller saturation value than $S_N$, but jumps are less prominent, and no banding is visible in any of the three plots.
It is clear that the phenomenology of entropy growth in the Schwinger model is distinct from that of ``typical'' random uniform disorder MBL.

However, there is a possibility that the observed behavior results from the discrete nature of the disorder.
To that end, we also look at a version of the XXZ model where $W_j$ instead takes on the discrete values $\{-W, 0, +W\}$.
If we choose $W = 20$ to be an even multiple of $J$, we can expect some  arguments of our discussion of fragmentation in the main text would continue to hold.
In Fig.~\ref{fig:app_xxz_w20}(g-i), we show $S_E$, $S_N$, and $S_C$ for this model, for $W = 20$.
This restores some of the phenomenology we observe in the Schwinger model, including slow growth in both $S_E$ and $S_C$, and some banding in all three types of entropy.
\end{document}